\documentclass[12pt]{article}
\usepackage{latexsym}
\usepackage[numbers]{natbib}
\usepackage{graphicx}
\usepackage{epsfig}
\usepackage{amssymb}
\usepackage{amsmath}
\usepackage{epstopdf}
\usepackage{eufrak}
\usepackage{epstopdf}
\usepackage{pb-diagram}
\usepackage{astrojournals}
\usepackage{url}
\usepackage[hmargin=1in,vmargin=1in]{geometry}

\usepackage[usenames, dvipsnames]{color}

\newcommand\Beq{\begin{eqnarray}} 
\newcommand\Eeq{\end{eqnarray}}

\newcommand{\eq}[1]{equation~(\ref{#1})}
\newcommand{\eqs}[2]{equations~(\ref{#1})~\&~(\ref{#2})}
\newcommand{\eqss}[2]{equations~(\ref{#1})--(\ref{#2})}

\newcommand{\eg}{\textit{e.g.}, }

\renewcommand{\vec}[1]{\boldsymbol{#1}}

\newcommand{\cross}{\vec{\times}}

\newcommand{\grad}{\vec{\nabla}}

\newcommand{\curl}{ \grad \cross }

\renewcommand{\l}{\ell}

\usepackage{changes}
\definecolor{DarkGreen}{rgb}{0,.55,0}
\definecolor{cerulean}{rgb}{0.0, 0.48, 0.65}
\definecolor{ultramarine}{rgb}{0, 0.125, 0.376}
\definechangesauthor[name={Sheehan}, color=DarkGreen]{so}
\definechangesauthor[name={Keaton}, color=cerulean]{kb}
\definechangesauthor[name={Oishi}, color=ultramarine]{jso}

\usepackage{todonotes}

\usepackage{verbatim}

\begin{document}

\renewcommand*{\thefootnote}{\fnsymbol{footnote}}

\centerline{\textbf{\large{Tensor calculus in spherical coordinates using Jacobi polynomials}}}
\vspace{0.2cm}
\centerline{\textbf{\large{Part-II: Implementation and Examples}}}
\medskip
\centerline{Daniel Lecoanet*$^{1,2}$, Geoff Vasil*$^3$, Keaton Burns$^4$, Ben Brown$^5$, Jeff Oishi$^6$}
\centerline{\textsl{\small{*Corresponding authors; email: lecoanet@princeton.edu}}}
\centerline{\textsl{\small{$^1$Princeton Center for Theoretical Science, Princeton, NJ 08544, USA}}}
\centerline{\textsl{\small{$^2$Princeton University Department of Astrophysical Sciences, Princeton, NJ 08544, USA}}}
\centerline{\textsl{\small{$^3$University of Sydney School of Mathematics and Statistics, Sydney, NSW 2006, Australia}}}
\centerline{\textsl{\small{$^4$Massachusetts Institute of Technology Department of Physics, Cambridge, MA 02139, USA}}}
\centerline{\textsl{\small{$^5$University of Colorado Laboratory for Atmospheric and Space Physics and Department of}}}
\centerline{\textsl{\small{ Astrophysical and Planetary Sciences, Boulder, CO 80309, USA}}}
\centerline{\textsl{\small{$^6$Bates College Department of Physics and Astronomy, Lewiston, ME 04240, USA}}}

\renewcommand*{\thefootnote}{\arabic{footnote}}

\bigskip
\centerline{\today}
\bigskip

\bigskip
\textbf{Abstract}
We present a simulation code which can solve broad ranges of partial differential equations in a full sphere. The code expands tensorial variables in a spectral series of spin-weighted spherical harmonics in the angular directions and a scaled Jacobi polynomial basis in the radial direction, as described in \citep{Vasil2018} (Part-I). Nonlinear terms are calculated by transforming from the coefficients in the spectral series to the value of each quantity on the physical grid, where it is easy to calculate products and perform other local operations. The expansion makes it straightforward to solve equations in tensor form (i.e., without decomposition into scalars). We propose and study several unit tests which demonstrate the code can accurately solve linear problems, implement boundary conditions, and transform between spectral and physical space. We then run a series of benchmark problems proposed in \citep{Marti2014}, implementing the hydrodynamic and magnetohydrodynamic equations. We are able to calculate more accurate solutions than reported in \citep{Marti2014} by running at higher spatial resolution and using a higher-order timestepping scheme. We find the rotating convection and convective dynamo benchmark problems depend sensitively on details of timestepping and data analysis. We also demonstrate that in low resolution simulations of the dynamo problem, small changes in a numerical scheme can lead to large changes in the solution. To aid future comparison to these benchmarks, we include the source code used to generate the data, as well as the data and analysis scripts used to generate the figures.


\bigskip
\bigskip
\bigskip
\bigskip
\bigskip
\bigskip
\bigskip
\bigskip
\bigskip
\bigskip
\bigskip
\bigskip

\pagebreak

\bigskip
\section{Introduction}

Stars and planets are spherical to an excellent approximation. This makes spherical coordinates a natural choice for solving problems in astrophysical and geophysical fluid dynamics (e.g., \citep{Miesch2000,Glatzmaier2013}, but see \citep{Joggerst2014} for an alternative view). The spherical coordinates $(r,\theta,\phi)$ have two types of coordinates singularities: at $\theta=0, \pi$; and at $r=0$. Functions written in spherical coordinates must satisfy regularity conditions near these coordinate singularities \cite[e.g.,][]{Livermore2007}. 

In \citep[][hereafter, Part-I]{Vasil2018}, we discuss a strategy for computing general tensor-calculus operations on functions in the three-dimensional ball. This naturally leads to methods for solving a wide class of partial differential equations (PDEs) in spherical coordinates. We expand each of the PDEs' dependent variables in a spectral series using spin-weighted spherical harmonics for the $\theta$ and $\phi$ dependence \citep[e.g.,][]{Phinney1973} and a scaled class of Jacobi polynomials for the $r$ dependence \citep[similar to][]{Matsushima1995}. Each basis function satisfies the regularity conditions at the coordinate singularities, so their sum automatically does as well.  This is a similar to our approach to simulations in cylindrical geometry \citep{Vasil2016}. The results from the disk provide an introduction to the more complex geometry of the full three-dimensional ball.

Previous researchers \citep[e.g.,][]{Livermore2007, Matsushima1995} have derived similar radial basis functions for scalar variables. Part-I provides a more thorough overview of the numerous different methods that have been developed to accurately cope with the large dynamic range associated with polar coordinate singularities.  
\citep{Boyd2011} also provide an excellent introduction to the topic in general.  Although vectors and higher order tensors can be decomposed into their scalar components (e.g., toroidal--poloidal decomposition for divergence-free vectors), this becomes tedious for high rank tensors. In contrast, Part-I derives basis functions for arbitrary tensorial variables. Tensors of different ranks are linked by sparse derivative operators, expressing various tensorial relations. For example, the gradient of a vector is a rank-2 tensor, the divergence of a vector is a scalar, etc. This makes it possible to solve tensorial equations in primitive form (e.g., without decomposition into scalars), making this method applicable to wide classes of PDEs.  \citep{James1976} derived a similar basis for tensorial quantities, but mostly focused on the cartesian components (i.e., $x$, $y$, $z$ components of a tensor).

This paper contains a series of tests which demonstrate the utility of our method.  We run unit tests: eigenvalue problems (section~\ref{sec:EVP}) and boundary value problems (section~\ref{sec:BVP}). These tests demonstrate that we can accurately solve linear problems and perform transforms from physical space to spectral space. They also test our implementation of a wide variety of boundary conditions that are used in hydrodynamics and magnetohydrodynamics.

We also run full-code tests. In section~\ref{sec:IVP} we simulate all three of the full-sphere benchmark problems described in \cite[][hereafter, M14]{Marti2014}. Section 6 of this paper can be thought of as a follow-up to M14 as we calculate converged solutions to higher precision by running at higher resolution and with higher-order time steppers. We include details of our simulations and data analysis that are necessary to compare between codes (e.g., timestepping scheme, timestep size, etc.). The supplementary materials also include the full source code so the interested reader could confirm any details of the simulations. Reduced data outputs and analysis scripts are public and in the repository \url{https://github.com/lecoanet/dedalus_sphere} and at \url{https://princeton.edu/~lecoanet/data}.

\section{Summary of the Algorithm}
In this paper we solve initial value problems, boundary value problems, and eigenvalue problems using the algorithms derived in Part-I.  As an example, consider the initial value problem,
\Beq\label{eqn:IVP}
M.\partial_t X + L. X = F(X),
\Eeq
where $X$ is a state vector consisting of a list of tensorial fields.  Examples from fluid dynamics include: scalar fields (rank 0), e.g., density, temperature, pressure, the divergence of the velocity; vector fields (rank 1), e.g., velocities, magnetic fields, temperature gradient; rank 2 tensor fields, e.g., the strain rate, Maxwell stress; and higher order tensors.  $M$ and $L$ are linear operators, possibly including derivative operators such as gradients, divergence, curl, etc.  $F$ is a general nonlinear function.

We solve \eq{eqn:IVP} in spherical polar coordinates $(r,\theta,\phi)$, with $r\in [0,1]$, $\theta\in [0,\pi]$, and $\phi\in [0, 2\pi)$, subject to the boundary conditions
\Beq
B.X|_{r=1} = E(X|_{r=1}),
\Eeq
where $B$ is a linear operator and $E$ is a nonlinear function, and initial conditions
\Beq
X|_{t=0}=X_0.
\Eeq
There are analogous formulations of boundary value and eigenvalues problems that we  discuss below.

Our approach is to expand $X$ in the bases described in Part-I, and then rewrite the problem in terms of the coefficients of the basis elements. Here we briefly summarize some important results of Part-I.  Consider a rank-$\mathfrak{r}$ tensor $\mathrm{T}$.  Then $\mathrm{T}$ has $3^\mathfrak{r}$ components, corresponding to a linear combination of tensor products of the coordinate unit vectors $\vec{e}_r$, $\vec{e}_\theta$, and $\vec{e}_\phi$.  The element $e(i)$ represents a single tensorial component using multi-index notation. For example for rank-3 tensors, $i = \{0,0,1\}$ corresponds to $e(i) = \vec{e}_{r} \otimes \vec{e}_{r} \otimes \vec{e}_{\theta}$. See the appendix in Part-I for a discussion of the multi-index notation. Therefore, 
\Beq\label{eqn:expansion}
\mathrm{T}(r,\theta,\phi) = \sum_{i,\sigma,\mathrm{a}} \sum_{m,\ell} \sum_{n} \, \hat{T}^{\mathrm{a}}_{m,\ell,n} \, Q^{\alpha,\ell+\bar{\mathrm{a}}}_{n}(r) \, \mathcal{Q}_\ell(\sigma,\mathrm{a}) \, Y_{\ell,m}^{\bar{\sigma}}(\theta,\phi) \, U(\sigma,i) \, e(i),
\Eeq
where the $Q_{n}^{\alpha,\l+\overline{\mathrm{a}}}(r)$ are related to a set of Jacobi polynomials, $\mathcal{Q}_{\l}$ is an $\ell$-dependent orthogonal transformation, $Y_{\ell,m}^{\bar{\sigma}}(\theta,\phi)$ are the spin-weighted spherical harmonics, $U(\sigma,i)$ is a unitary transformation, and overbars denote a sum over multi-indices.  Thus, $\hat{T}^{\mathrm{a}}_{m,\ell,n}$ is a (complex-valued) coefficient of $\mathrm{T}$ using this basis.  Our choice of basis ensures the solutions satisfy regularity conditions at the poles ($\theta=0$ and $\pi$) and the origin ($r=0$), and these bases ensure that derivative operators are maximally sparse (Part-I).

For calculations, we must truncate the sums in \eq{eqn:expansion}.  We pick a value of $L_{\rm max}$ and $N_{\max}$. The maximum spherical harmonic degree is $L_{\rm max}$, and the azimuthal order $m$ ranges $0\leq m \leq L_{\rm max}$. Note that we do not need negative values of $m$ because we assume the tensor field is real. The truncation for $\ell$ modes depends on the spin of the component in question, $\sigma$. For each azimuthal order $m$, we have $\max(m,|\bar{\sigma}|)\leq \ell \leq L_{\rm max}$. We use $\bar{\cdot}$ to denote the sum of the elements of a given spin or regularity multi-index, i.e., if $\sigma=\{+1,-1\}$, then $\bar{\sigma}= (+1) + (-1) =  0$ (see Part-I for more details). The truncation for tensors is similar to the familiar triangular truncation for scalar spherical harmonics. The truncation requires that the degree of $\sin(\theta)$ in the spin-weighted spherical harmonic is no greater than $L_{\rm max}$.

The $Q$ polynomials are
\Beq
Q^{\alpha,\ell+\bar{\mathrm{a}}}_{n}(r) \ \propto \ r^{\ell+\bar{\mathrm{a}}} P^{(\alpha,\ell+\bar{\mathrm{a}}+1/2)  }_{n} (2r^2-1),
\Eeq
where $P_n$ is a Jacobi polynomial of degree $n$ and the proportionality is determined by a normalization factor.  Thus, this $Q$ polynomial has an $r$ degree of $\ell+\bar{\mathrm{a}}+2n$.  Similar to spherical harmonics, we also impose a triangular truncation, requiring that the radial degree of each $Q$ polynomial is bounded. For each problem, we determine the highest tensor rank we are interested in, $\mathcal{R}_{\rm max}$, so $\bar{\mathrm{a}}\leq \mathcal{R}_{\rm max}$. We then require that $ 0\leq 2 n \leq 2 N_{\rm max} - \ell + \mathcal{R}_{\rm max}$. The range of $n$'s depends on the value of $\ell$. Our use of $\mathcal{R}_{\rm max}$ means there are the same number of $n$ for tensor components with different $\mathrm{a}$, which greatly simplifies our analysis.

The sum over $i$, $\sigma$, and $\tau$ are, respectively, sums over the spherical components of the tensor $\mathrm{T}$, the spin indices, and the regularity indices.  Each multi-index of a rank-$\mathfrak{r}$ tensor has $3^\mathfrak{r}$ elements.  There is also one additional index, $\alpha$. All fields start off with $\alpha=0$, but operations like differentiation increase the value of $\alpha$ by one. We use different values of $\alpha$ to keep the differentiation matrices maximally sparse. This is equivalent to the sparse derivative relation between Chebyshev-$T$ and Chebyshev-$U$ polynomials. We use conversion matrices to ensure all variables in an equation have the same value of $\alpha$. 

After the truncation of \eq{eqn:expansion}, the state vector consists of $\mathcal{O}(L_{\rm max}^2N_{\rm max})$ expansion coefficients for each tensor component.  In this paper we only study problems in which the linear operators (e.g., $M$, $L$, $B$) do not contain any explicit dependence on $\theta$ or $\phi$ (but we allow coupling in all directions through  gradient operators).  In this case, the linear operators only couple different radial modes together.  Thus, we can consider \eq{eqn:IVP} as $\mathcal{O}(L_{\rm max}^2)$ different equations for $X_{m,\ell}$, the state vector corresponding to spherical harmonic order $m$ and $\ell$.  Each tensor component has $\mathcal{O}(N_{\rm max})$ components in $X_{m,\ell}$.  The linear operators acting on $X_{m,\ell}$ are sparse with $\mathcal{O}(N_{\rm max})$ elements.  They can be easily inverted with off-the-shelf sparse linear algebra packages.  Coupling in only the radial direction and between different field variables allows the parallelization of the code across both $m$ and $\ell$.  With this restriction on linear operators, we cannot treat terms like the Coriolis force with implicit time stepping (though we can treat it as a part of the nonlinear operator $F(X)$).

To calculate the nonlinear terms $F(X)$ and $E(X)|_{r=1}$, we transform the coefficients $\hat{T}_{m,\ell,n}^{\mathrm{a}}$ into the tensor field $\mathrm{T}(r,\theta,\phi)$ in physical space, and then perform any local operations (e.g., products) in physical space.  The transform requires $\mathcal{O}(L_{\rm max}^2)$ matrix-multiply transforms for the radial basis (i.e., multiplication by a dense, $\mathcal{O}(N_{\rm max}^2)$ matrix), and $\mathcal{O}(L_{\rm max}N_{\rm max})$ matrix-multiply transforms for the angular basis (i.e., multiplication by a dense, $\mathcal{O}(L_{\rm max}^2)$ matrix).  Thus, the transformations require $\mathcal{O}(N_{\rm max}L_{\rm max}^2 \max(N_{\rm max},L_{\rm max}))$ operations and are expected to be the slowest part of the calculation when $L_{\rm max}$ and $N_{\rm max}$ become large.  Practically speaking, these transformations are reliant on the speed of the linear algebra library, in particular the speed of matrix and vector dot products; these are typically well-optimized numerical operations.

In section~\ref{sec:transforms}, we describe how we transform between data in physical space ($\mathrm{T}(r,\theta,\phi)$), and the coefficient expansion of \eq{eqn:expansion} ($\hat{T}_{m,\ell,n}^{\mathrm{a}}$).  In the subsequent sections, we  describe the implementation of this formulation for eigenvalue problems (section~\ref{sec:EVP}), a boundary value problem (section~\ref{sec:BVP}), and the three initial value problems described in M14 (section~\ref{sec:IVP}).

\section{Transforms}\label{sec:transforms}

Here we describe how we transform data from physical space ($\mathrm{T}(\phi,\theta,r)$), back and forth from the coefficient expansion ($\hat{T}_{m,\ell,n}^{\mathrm{a}}$) in terms of scaled Jacobi polynomials in the radial direction and spin-weighted spherical harmonics in the angular directions.  This is crucial for efficiently calculating nonlinear products, as well as visualizing our data.

Consider a rank-$\mathfrak{r}$ tensor $\mathrm{T}(\phi,\theta,r)$.  We represent $\mathrm{T}$ with $N_c = 3^\mathfrak{r}$ components, each of which have $N_r$ radial points, $N_\theta$ latitudinal points and $N_\phi$ longitudinal points.  The data are initially on the quadrature nodes of the spin-weighted spherical harmonics in $\phi$, $\theta$ and the Jacobi polynomials in $r$.  We assume the components of $\mathrm{T}$ are real.  We represent each component of $\mathrm{T}$ as a \verb!Field! in \verb!Dedalus!\footnote{More information and source code at \url{http://dedalus-project.org}.}, as it wraps \verb!FFTW!'s Fourier transforms and parallel transposes.  The data for $\mathrm{T}(\phi,\theta,r)$ are stored as a $N_c\times N_\phi\times N_\theta \times N_r$ \verb!numpy! array.

The grid points and transform matrices are calculated using Gaussian quadrature. We generate a guess for the quadrature grid and weights using the Golub-Welsch algorithm. After this we polish the results using a Newton iteration. We use Legendre quadrature for the latitudinal direction. We use Jacobi quadrature with parameters $(0,1/2)$ for the radial direction. After obtaining the grid, we compute all higher-order spin-weighted spherical harmonics and generalised spherical Zernike polynomials out of the three-term recursion for Jacobi polynomials. Each basis function comprises a spatial envelope (\eg $r^{\l}$) multiplied by a Jacobi polynomial of some kind (\eg $P_{n}^{(0,\l+1/2)}(2r^{2}-1)$ ) We initialise with the appropriate spatial envelope and recurse up to the desired degree from there. In some cases the spatial envelope contains an extreme dynamic range (\eg $r^{\l}$ for $\l \gg 1$). This can lead the initialization to underflow to zero. Grid points where this occurs can never return to finite values, even though the eventual basis function should end up moderate at such points. To avoid underflow problems, we used 128-bit precision for the construction phase of the transform grid, weights and matrices. We cast the results to 64-bit precision after the initial construction. There are more sophisticated methods available to handle the same problems, e.g., the recursions presented in \citep{Risbo1996}, but our simple method works for all polynomial degrees up to roughly $\approx 10,000$. The speed of the construction is fast enough considering we store the transform matrices for later use.  

Although some of our simulations use dealiasing, others do not. The simulations which were dealiased were neither systematically less accurate, nor more accurate, than the simulations without dealiasing. Simulations without dealiasing have $N_\phi=2(L_{\rm max}+1)$, $N_\theta=L_{\rm max}+1$, and $N_r=N_{\rm max}+1$. Simulations with dealiasing have $N_\phi=3(L_{\rm max}+1)$, $N_\theta=\frac{3}{2}(L_{\rm max}+1)$, and $N_r=\frac{3}{2}(N_{\rm max}+1)$.

We can transform data from physical space to coefficient space, and back, either in serial or parallelized across  cores (using MPI).  We can parallelize in either one or two directions for 3D calculations.  If the data size is $\mathcal{O}(N)$ in each direction, parallelization across two directions allows a calculation to be run efficiently on $\mathcal{O}(N^2)$ cores.  We  describe the transform assuming parallelization across two directions, but  note how the calculation differs if parallelized in a single direction.

In physical space, each core has the data for all $\phi$ points, but for only a subset of points in $\theta$ and $r$ (or $r$ only for parallelization across one direction).  We say that the data are local in $\phi$, but distributed across $\theta$ and $r$.  First we use \verb!Dedalus! to perform a real to complex Fourier transform in $\phi$, so we have $\mathrm{T}_m(\theta,r)$.    If the data is distributed across cores in $\theta$, we use \verb!Dedalus! to perform a parallel transpose across $m$ and $\theta$ so the data are local in $\theta$ and distributed across $m$.  Each processor has data for a sequential subset of $m$ values.

Next we multiply by the unitary matrix $U^\dagger$ to transform from the components $i$ to spins $\sigma$.  For each $m$ and spin $\sigma$, we use a matrix multiplication transform to calculate the spin-weighted spherical harmonic coefficients
\Beq
\hat{T}^\sigma_{m,\ell}(r) = \sum_{i} S^{\bar{\sigma}}_{\l}(\theta_{i})\, \hat{T}^\sigma_{m}(\theta_{i},r),
\Eeq
where the matrices $S^{\bar{\sigma}}_{\l}(\theta_{i})$ represent whichever spin-weighted transform matrix is appropriate at the time; each matrix has size $(L_{\rm max}-L_{\rm min}+1)\times N_\theta$, where $L_{\rm min}=\max(|m|,|\sigma|)$.  Although there is less data for higher $m$, we pad with zeros so $\hat{T}^\sigma_{m,\ell}(r)$ is a rectangular $N_c\times (L_{\rm max}+1)\times (L_{\rm max}+1) \times N_r$ array. The $S$ matrix is the product of a spin-weighted spherical harmonic function and Gaussian quadrature weights. The inverse transform matrix is simply a spin-weighted spherical harmonic function, properly transposed \citep{Boyd_book}.

We next use \verb!Dedalus! to perform a parallel transpose across $\ell$ and $r$ so the data are local in $r$ and distributed across $m$ and $\ell$ (or $\ell$ only if parallelized across one direction).  Each processor has data for a sequential subset of the $\ell$ values.

We next multiply by the $\ell$-dependent orthogonal matrix $\mathcal{Q}_\ell^{\top}$ to transform from the spins $\sigma$ to regularities $\mathrm{a}$.  This gives $\hat{T}^{\mathrm{a}}_{m,\ell}(r)$.  Then for each $\ell$ we use a matrix multiplication transform to calculate the Jacobi polynomial coefficients,
\Beq
\hat{T}_{m,\ell,n}^{\mathrm{a}} =  \sum_{i} w(r_{i})\, {Q}_{n}^{\l+\bar{\mathrm{a}}}(r_{i}) \, \hat{T}^{\mathrm{a}}_{m,\ell}(r_{i}),
\Eeq
where $w(r_{i})$ is the Gaussian quadrature weight, and ${Q}_{n}^{\l+\bar{\mathrm{a}}}(r_{i})$ represents whichever Jacobi-polynomial transform matrix is appropriate at the time; each matrix has size $(N_{\rm max}-N_{\rm min}+1)\times N_r$, where $N_{\rm min}=\lfloor|\ell-\mathcal{R}_{\rm max}|/2\rfloor$.  The data are then stored as a list of arrays with $N_C(N_{\rm max}-N_{\rm min}+1) \times N_m$ elements for each value of $\ell$. The forward transform is the product of a Zernike polynomial and Gaussian quadrature weight. The inverse radial transform matrix is simply a Zernike polynomial, properly transposed.

This gives the coefficient expansion, $\hat{T}_{m,\ell,n}^{\mathrm{a}}$ associated with the tensor $\mathrm{T}(r,\theta,\phi)$.  As each step is invertible, the algorithm can be inverted to calculate the grid values $\mathrm{T}(r,\theta,\phi)$ which correspond to coefficients $\hat{T}_{m,\ell,n}^{\mathrm{a}}$.

\section{Eigenvalue Problems}\label{sec:EVP}

In this section, we  solve eigenvalue problems of the form,
\Beq\label{eqn:EVP}
\lambda M.X + L.X=0,
\Eeq
subject to boundary conditions
\Beq
B.X|_{r=1} = 0,
\Eeq
where $M$, $L$, and $B$ are linear operators, and $\lambda$ is the eigenvalue.

\subsection{Surface Rossby Waves}\label{sec:rossby}

Before discussing problems in the full sphere, we  briefly mention an example of two-dimensional flow on the surface of a sphere. In this case, we expand the solution only in spin-weighted spherical harmonics. We  consider Rossby waves. We solve the Laplace-tidal equations
\Beq
-i\omega \,\vec{u} + \vec{\nabla} p + \cos(\theta)\vec{e}_r\vec{\times}\vec{u} &=& 0, \\
-i\omega \,\gamma\, p + \vec{\nabla}\vec{\cdot}\vec{u} &=& 0,
\Eeq
on the sphere $r=1$, where we normalize the problem such the Coriolis parameter is equal to unity. Lamb's parameter $\gamma \ = \  4 \Omega^{2} a^{2}/gH$, which we take to be zero. The statevector is
\Beq
X \ = \ \left[
\begin{array}{c}
 u^- \\
 u^+ \\
 p
\end{array}
\right],
\Eeq
where $u^-$ and $u^+$ are the two spin components of the angular velocities.  The linear operators $M$ and $L$ are
\Beq
M \ = \ \left[
\begin{array}{ccc}
 i & 0 & 0 \\
 0 & i & 0 \\
 0 & 0 & i \,\gamma 
\end{array}
\right], \quad\quad\quad\quad
L \ = \ \left[
\begin{array}{ccc}
 -i\mathcal{C} & 0 & k^-  \\
 0 & i\mathcal{C} & k^+ \\
 k^+ & k^- & 0 
\end{array}
\right],
\Eeq
where again we take $\gamma=0$.

\begin{figure}
  \centerline{\includegraphics[width=7.1in]{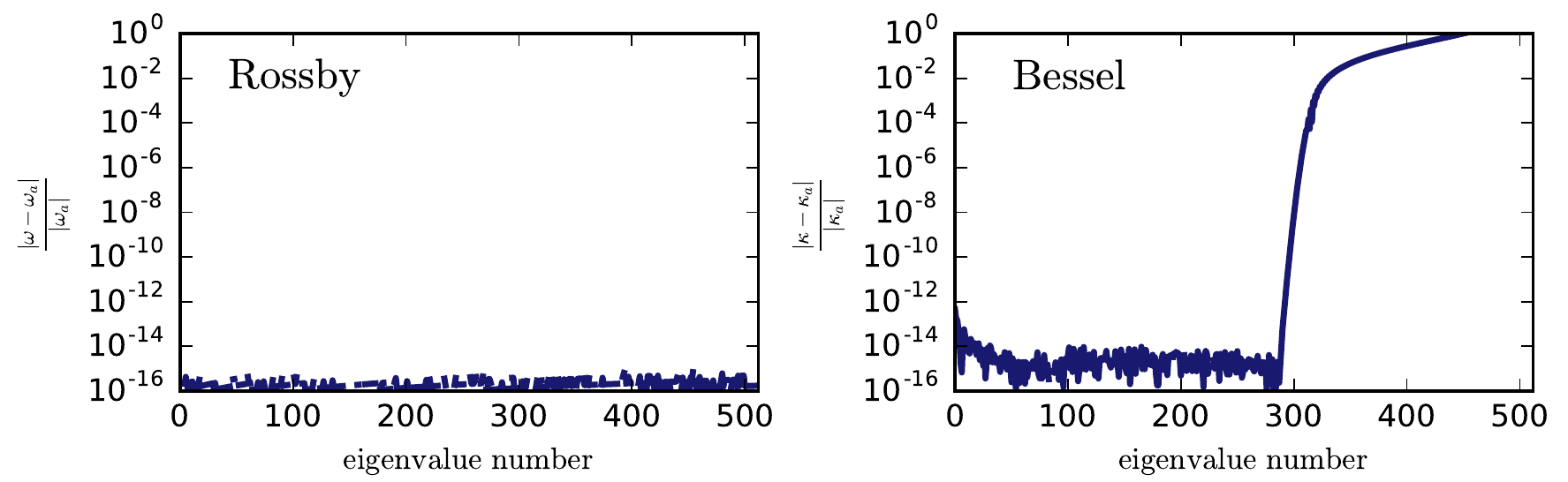}}
  \caption{The fractional error in the eigenvalues in the spherical Rossby wave problem (left panel) and the Bessel function problem (right panel).  For the Rossby wave problem, machine precision is achieved for all eigenvalues because the eigenfunctions are closely related to our basis functions.  For the Bessel function problem, the eigenfunctions are not the same as the radial basis functions, and owing to this, roughly half of the eigenvalues are at machine precision while the other half have large errors.}
\label{fig:eigenvalues}
\end{figure}

We expand $p$, and $u^\pm$ in 512 spin-weighted spherical harmonics each.  See Part-I for the definitions of the cosine operators $\mathcal{C}$ and the angular derivatives $k^\pm$. The problem is coupled in $\ell$ but not in $m$. Thus, we can solve each $m$ independently.  We pick $m=50$ for illustrative purposes. The generalized eigenvalue problem is solved using \verb!scipy!'s \verb!eig! routine. The analytical eigenvalues are
\Beq
\omega_a = - \frac{m}{\ell(\ell+1)}.
\Eeq
This is a non-trivial problem because eigen-solutions cannot be represented in terms of a pure spin-weighted spherical harmonic; multiplication by $\cos(\theta)$ complicates the situation. However, the solution is expressible in a simple finite combination of spherical harmonics. We demonstrate in figure~\ref{fig:eigenvalues} that we correctly find all the eigenvalues.

\subsection{Spherical Bessel Equation}\label{sec:bessel}

We next solve the  scalar-Laplacian eigenvalue equation  
\Beq\label{eqn:bessel}
\nabla^2 f + \kappa^2 f = 0,
\Eeq
with the boundary condition,
\Beq
f(r=1) = 0.
\Eeq
This is an eigenvalue problem with eigenvalue $\kappa^2$.  We posed this problem as an example of our approach in Part-I, here we provide the full numerical solution.  The equation is separable into a radial and an angular component.  The scalar function $f$ can be expanded in scalar spherical harmonics, $R_{\ell,m}(r)Y^0_{\ell,m}(\theta,\phi)$.  Then $R_{\ell,m}$ satisfies the spherical Bessel equation
\Beq
r^2 \frac{d^2 R}{dr^2}+2r \frac{dR}{dr} + \left(\kappa^2 r^2 - \ell (\ell+1)\right) R = 0.
\Eeq
The solutions are spherical Bessel functions of the first kind, $j_{\ell}(\kappa r)$.  The boundary condition at $r=1$ requires $\kappa$ to be a zero, i.e., $j_{\ell}(\kappa)=0$.

To solve this numerically, we take the state vector $X=f$, $L=D_{1,\ell+1}^{-}D_{0,\ell}^{+}$, and the eigenvalue $\lambda=\kappa^2$.  Because $L.X$ is in the $\alpha=2$ function space, owing to the two derivative operators, in equation~(\ref{eqn:EVP}) the matrix $M=C_{1,\ell} C_{0,\ell}$, where the $C_{\alpha,\ell}$ are conversion matrices which increment $\alpha \to \alpha+1$. Paper-I discusses the details of how we construct a matrix system of equations from the original PDE~(\ref{eqn:bessel})

We expand $f$ in a basis of 512 polynomials.  To implement the boundary condition, the last row of $L$ is replaced with the $r=1$ restriction operator $Q^{\alpha,\ell}(r=1)$, a row vector of the $Q$ polynomials evaluated at $r=1$, and the last row of $M$ is replaced by zeros.  The generalized eigenvalue problem is solved using \verb!scipy!'s \verb!eig! routine.  The eigenvectors are transformed to the grid to compare to the spherical Bessel function.  

\begin{figure}
  \centerline{\includegraphics[width=7.1in]{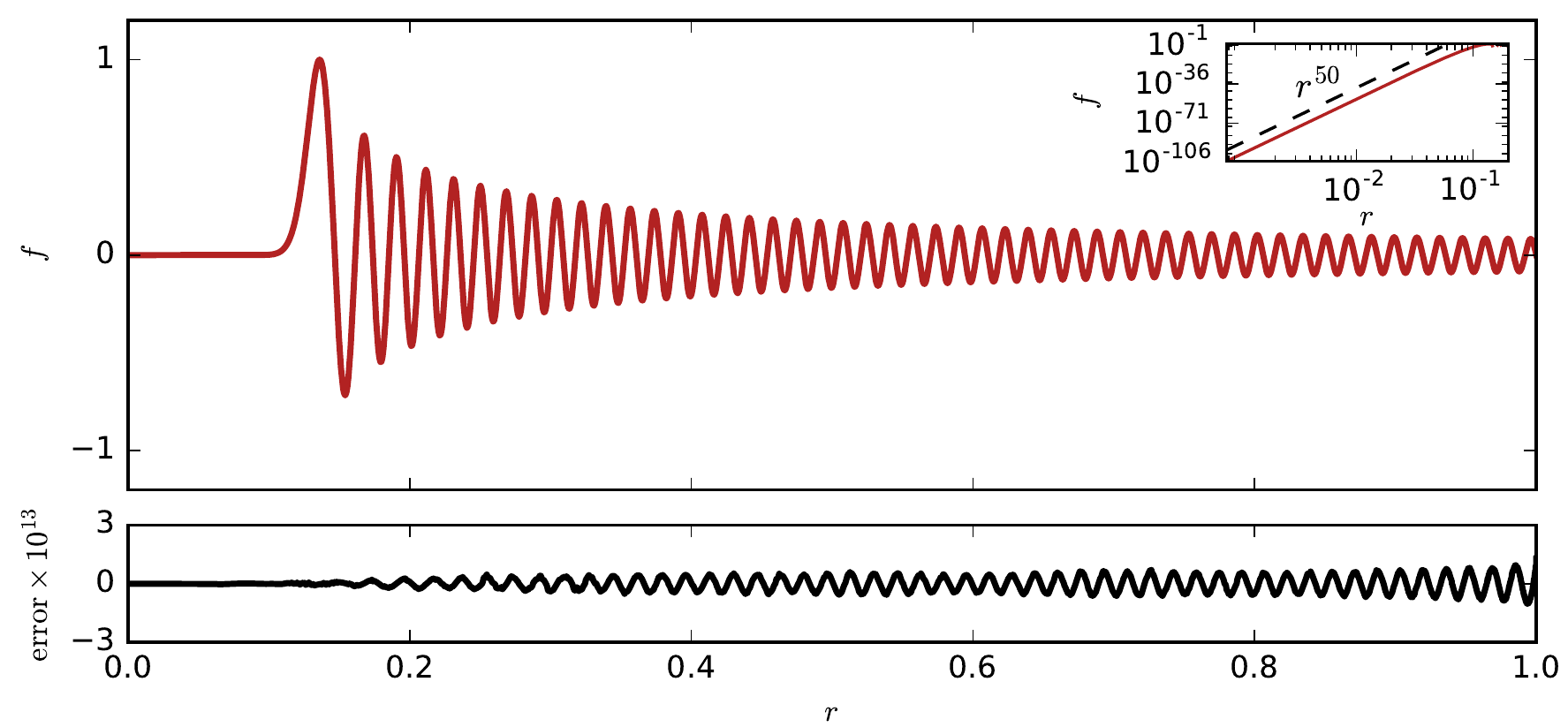}}
  \caption{Top panel: The 100$^{\rm th}$ eigenmode solution to the spherical Bessel equation with $\ell=50$.  The inset shows the solution has the correct power-law behavior as $r$ approaches zero.  Bottom panel: The error of the solution, $f-j_{\ell}(\kappa r)$ stays very small throughout the entire domain.}
\label{fig:bessel eigenfunction}
\end{figure}

In figure~\ref{fig:bessel eigenfunction}, we plot the 100$^{\rm th}$ eigenmode solution to \eq{eqn:bessel} with $\ell=50$ (top panel), along with the error ($|f-j_{\ell}(\kappa r)|$; bottom panel).  The inset shows that near the origin, $f\sim r^{50}$, as required by the regularity condition at $r=0$.  This regularity condition is satisfied automatically by our choice of radial basis.  Figure~\ref{fig:eigenvalues} plots the fractional error in the eigenvalue $\kappa$, where the analytic eigenvalues are the zeros $j_\ell(\kappa_a)=0$. As expected, about the first half of the eigenvalues are very accurate.  Eigenvalues corresponding to eigenmodes with high radial wavenumbers tend to have higher errors; these eigenvalues can be computed to machine precision by increasing the number of polynomials in the basis.

\subsection{Linear Diffusion Equation \& Boundary Conditions}\label{sec:diffusion}

In this section, we  solve the linearized, diffusive hydrodynamics equations
\Beq\label{eqn:diffusion_vel}
\partial_t \vec{u} + \vec{\nabla}p - \nabla^2\vec{u} &=& 0, \\
\vec{\nabla}\vec{\cdot}\vec{u} &=& 0.\label{eqn:diffusion_div}
\Eeq
We assume $\partial_t\vec{u}=-\kappa^2\vec{u}$, and solve for the damping rate $\kappa^2$.  Here we denote the pressure with $p$.  This is completely equivalent to the linearized equation for the magnetic vector potential $\vec{A}$ in the Coulomb gauge (i.e., $\vec{\nabla}\vec{\times}\vec{A}=\vec{B}$ for the magnetic field $\vec{B}$). The linearized induction equation is then
\Beq\label{eqn:diffusion_potential}
\partial_t \vec{A} + \vec{\nabla}\Phi - \nabla^2\vec{A} &=& 0, \\
\vec{\nabla}\vec{\cdot}\vec{A} & = &0.\label{eqn:diffusion_gauge}
\Eeq
Here $\Phi$ is the scalar potential. The general solution can be derived analytically, so we can implement a wide range of boundary conditions and calculate exact solutions.  This makes this problem very useful for insuring the proper implementation of boundary conditions in the code.

The hydrodynamics problem can have no-slip boundaries, or stress-free boundaries at $r=1$.
\Beq
\label{vec no-slip}
\vec{u} &=& 0, \quad \quad {\rm (no-slip)} \\
\vec{e}_r\vec{\cdot}\vec{u}=\vec{e}_\theta\vec{\cdot}{\mathrm E}\vec{\cdot}\vec{e}_r = \vec{e}_\phi\vec{\cdot}{\mathrm E}\vec{\cdot}\vec{e}_r &=& 0, \quad\quad {\rm (stress-free)}
\Eeq
where we have also assumed impenetrability and where
\Beq
{\mathrm E} = \frac{1}{2}\left(\vec{\nabla}\vec{u} + (\vec{\nabla}\vec{u})^{\top}\right),
\Eeq
is the rank-2 stress tensor.

There are also several choices for magnetic boundary conditions.  Potential boundary conditions assume the magnetic field matches onto a harmonic field for $r>1$. This is a non-local condition that is commonly specified by decomposing $\vec{A}$ into spherical harmonic degrees, $\vec{A}_\ell$. A perfectly-conducting boundary has no normal magnetic field and no tangential electric fields. The pseudo-vacuum boundary condition is that the tangential magnetic field is zero.
\Beq\label{eqn:potential}
\partial_r \vec{A}_\ell + (\ell+1)\vec{A}_\ell/r & = & 0, \quad\quad {\rm (potential)} \\
\vec{e}_\theta\vec{\cdot}\vec{A} = \vec{e}_\phi\vec{\cdot}\vec{A}=\Phi&=&0, \quad\quad {\rm (perfectly-conducting)} \\
\vec{\nabla}\vec{\cdot}\vec{A}=\vec{e}_\theta\vec{\cdot}\vec{\nabla}\vec{\times}\vec{A}=\vec{e}_\phi\vec{\cdot}\vec{\nabla}\vec{\times}\vec{A}&=&0. \quad\quad {\rm (pseudo-vacuum)} \label{eqn:pseudo-vacuum}
\Eeq

We solve each of these problems analytically in appendix~\ref{sec:analytic_solution}. In each case, the eigenvalues are related to the zeros of spherical Bessel functions of different orders.

In this problem the statevector is
\Beq
X \ = \ \left[
\begin{array}{c}
 u^- \\
 u^0 \\
 u^+ \\
 p
\end{array}
\right],
\Eeq
where $u^-$, $u^0$, and $u^+$ are the components of $\vec{u}$ in regularity classes.  The linear operators $M$ and $L$ are
\Beq
M \ &=& \ \left[
\begin{array}{cccc}
 C_{1,\ell-1}C_{0,\ell-1} & 0 & 0 & 0 \\
 0 & C_{1,\ell}C_{0,\ell} & 0 & 0 \\
 0 & 0 & C_{1,\ell+1}C_{0,\ell+1} & 0 \\
 0 & 0 & 0 & 0
\end{array}
\right], \\
L \ &=& \ \left[
\begin{array}{cccc}
 -D_{1,\ell}^-D_{0,\ell-1}^+ & 0 & 0 & \xi_\ell^-C_{1,\ell-1}D_{0,\ell}^- \\
 0 & -D_{1,\ell+1}^-D_{0,\ell}^+ & 0 & 0 \\
 0 & 0 & -D_{1,\ell}^+D_{0,\ell+1}^- & \xi_{\ell}^+C_{1,\ell+1}D_{0,\ell}^+ \\
 \xi_{\ell}^-D_{0,\ell-1}^+ & 0 & \xi_{\ell}^+D_{0,\ell+1}^- & 0
\end{array}
\right].
\Eeq
For no-slip boundary conditions, all three components of $\vec{u}$ are zero at the boundary.  For stress-free boundary conditions, to impose no normal flow, we set
\Beq\label{eqn:no normal flow}
\sum_{\mathrm{a}}\mathcal{Q}_\ell(0,\mathrm{a})\, Q^{\alpha,\ell+\mathrm{a}}(r=1) \cdot  u_\ell^\mathrm{a} = 0,
\Eeq
where $\mathcal{Q}_{\ell}$ is the rank one orthogonal matrix and $Q^{\alpha,\ell+\mathrm{a}}(r=1)$, the restriction operator, is a row vector of the $Q$ polynomials evaluated at $r=1$. In \eq{eqn:no normal flow}, $\alpha=0$.  The two other conditions are equivalent to
\Beq
\vec{e}_+\vec{\cdot} {\mathrm E} \vec{\cdot}\vec{e}_0 = \vec{e}_-\vec{\cdot} {\mathrm E} \vec{\cdot}\vec{e}_0.
\Eeq
Thus we impose
\Beq
\sum_{\mathrm{a},\mathrm{b}=-1}^{+1}\left[\mathcal{Q}_{\ell}(0+,\mathrm{a}\mathrm{b}) + \mathcal{Q}_\ell(+0,\mathrm{a}\mathrm{b})\right] \xi^\mathrm{a}_{\ell+\mathrm{b}} Q^{\alpha,\ell+\overline{\mathrm{a}\mathrm{b}}}(r=1) \cdot D_{0,\ell+\mathrm{b}}^{\mathrm{a}} u_\ell^\mathrm{b}&=&0, \\
\sum_{\mathrm{a},\mathrm{b}=-1}^{+1}\left[\mathcal{Q}_{\ell}(0-,\mathrm{a}\mathrm{b}) + \mathcal{Q}_\ell(-0,\mathrm{a}\mathrm{b})\right] \xi^\mathrm{a}_{\ell+\mathrm{b}} Q^{\alpha,\ell+\overline{\mathrm{a}\mathrm{b}}}(r=1)  \cdot D_{0,\ell+\mathrm{b}}^{\mathrm{a}} u_\ell^\mathrm{b}&=&0,
\Eeq
Note that here we must take $\alpha=1$ because the $D$ operator increases $\alpha$ from 0 to 1.

For the magnetic problem, the $M$ and $L$ matrices are identical, but the statevector changes to
\Beq
X \ = \ \left[
\begin{array}{c}
 A^- \\
 A^0 \\
 A^+ \\
 \Phi
\end{array}
\right].
\Eeq
The magnetic boundary conditions have a simple form in terms of regularities:
\Beq\label{eqn:potential BC}
A^-=D^-_{0,\ell+1}A^+=D^-_{0,\ell}A^0&=&0, \quad\quad {\rm (potential)} \\
A^0=\Phi=\xi^+_\ell A^--\xi^-_\ell A^+&=&0, \quad\quad {\rm (perfectly-conducting)} \\
A^-=D^-_{0,\ell+1}A^+=\left(D^-_{0,\ell}-\ell/r\right)A^0&=&0. \quad\quad {\rm (pseudo-vacuum)}
\Eeq
Although potential and pseudo-vacuum boundary conditions look similar, the pseudo-vacuum conditions can be expressed locally (\eq{eqn:pseudo-vacuum}), whereas the potential conditions cannot (\eq{eqn:potential}).

We apply boundary conditions using the tau method (see Part-I \& references within). Boundary conditions are imposed by adding a correction term (called $\tau$) to the equations. One arrives at (often subtly) different answers depending on the assumed form of the correction term.  We assume $\tau$ takes the form of one of our basis polynomials, $Q^{\alpha,\ell+\bar{\mathrm{a}}}_{n}(r)$, where $n$ is the highest order radial mode for the chosen values of $\ell$, $N_{\rm max}$, and $\mathcal{R}_{\rm max}$.

There is a remaining choice for what value of $\alpha$ to use.  We call this value $\alpha_{BC}$.  We use either $\alpha_{BC}=2$ or $\alpha_{BC}=0$.  Using $\alpha_{BC}=2$ is equivalent to replacing the last row of the $L$ matrix with the boundary condition, as we did in section~\ref{sec:bessel}.

For $\alpha_{BC}=0$, we add a single extra element to the state vector for each boundary condition, which corresponds to each $\tau$ correction.  Then we must add extra rows to the matrices, which are the boundary conditions.  We also must add extra columns to maintain square matrices.  The column associated with a given $\tau$ is given by the final column of the $C_{1,\ell}C_{0,\ell}$ matrix for the equation corresponding to the $\tau$ error (note it is $\ell$ dependent).  The extra columns are only non-zero for the variables for which we are applying boundary conditions, e.g., the divergence conditions are imposed exactly.  We have checked that if we instead use the final column of the identity matrix, we find the same results as for $\alpha_{BC}=2$.

To solve the problem numerically, we fix $\ell=50$ and use 256 terms in the radial expansion of each variable in the statevector.  We also apply boundary conditions using $\alpha_{BC}=2$ and $\alpha_{BC}=0$.  Thus, for $\alpha_{BC}=2$, the $M$ and $L$ matrices have size $1024^2$, whereas for $\alpha_{BC}=0$ they have size $1027^2$ because of the three extra rows and columns to incorporate the $\tau$ errors.  We solve this generalized eigenvalue problem with \verb!scipy!'s \verb!eig! routine.

\begin{figure}
  \centerline{\includegraphics[width=7.1in]{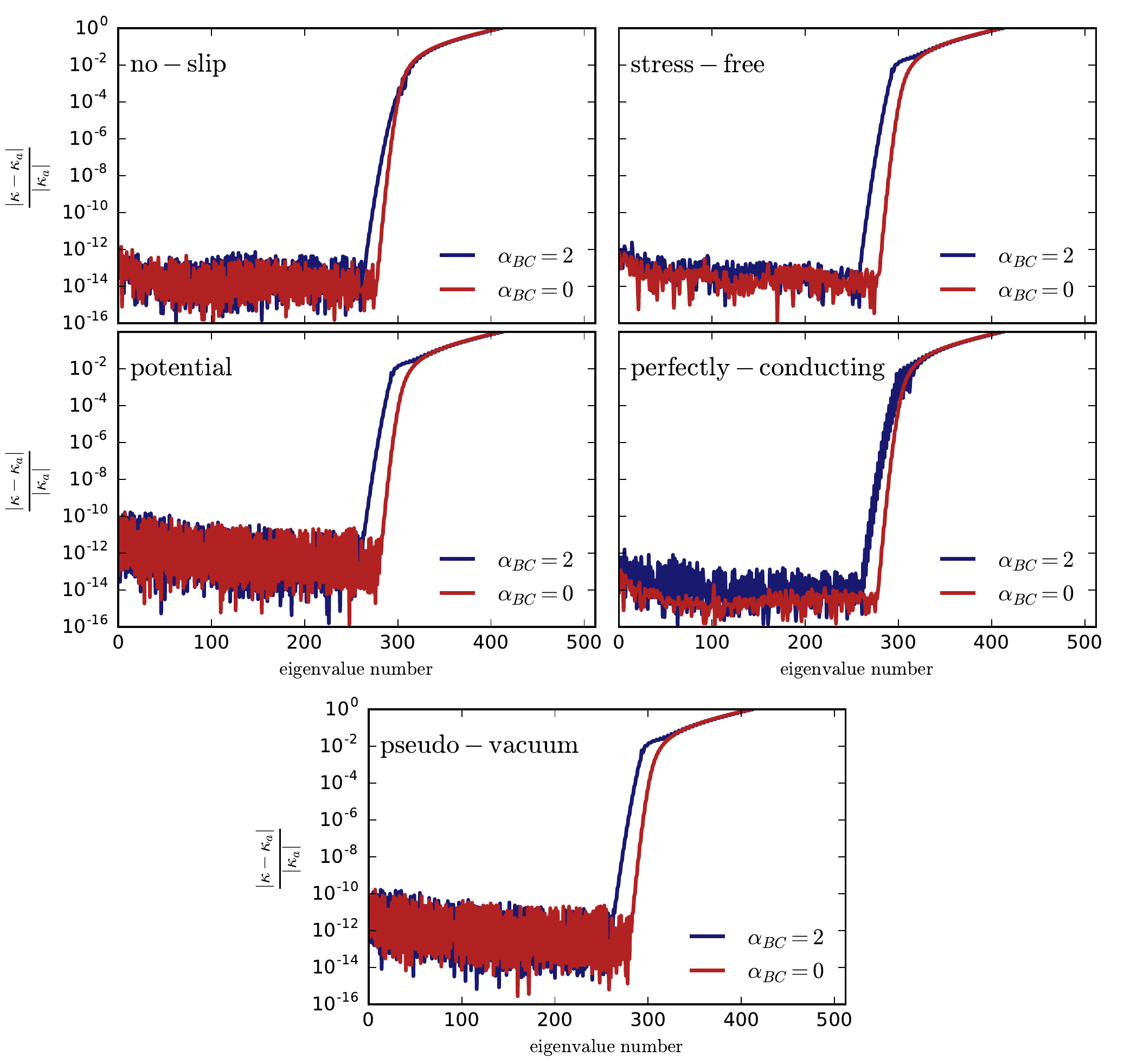}}
  \caption{The fractional error in the eigenvalue for the linear diffusion problem, with five different boundary conditions. In each case, we find that about half the eigenvalues are very accurate.  When we use $\alpha_{BC}=0$ to apply the boundary conditions, we find that the number of accurate eigenvalues is about  $10\%$ greater than when we use $\alpha_{BC}=2$.}
\label{fig:eigenvalues_BC}
\end{figure}

For each of these five sets of boundary conditions, we have two sets of eigenvalues because each problem decouples into a problem for the toroidal component, and a problem for the poloidal component (Appendix~\ref{sec:analytic_solution}).  We calculate the analytic values of the eigenvalues, $\kappa_a$, using the formulae in Appendix~\ref{sec:analytic_solution}. We sort the numerical and analytic eigenvalues and compare them.

Figure~\ref{fig:eigenvalues_BC} shows that about half of our eigenvalues are accurate.  The errors using $\alpha_{BC}=2$ and $\alpha_{BC}=0$ are similar for all boundary conditions except perfectly-conducting, where $\alpha_{BC}=0$ is more accurate.  However, for each case, when we use $\alpha_{BC}=0$ there are about $\sim 20$ extra accurate eigenvalues compared to $\alpha_{BC}=2$.  This suggests that using $\alpha_{BC}=0$, we are able to correctly resolve smaller scale features at a given resolution.  Note that since our resolution is 256, we expect to have around 512 eigenvalues because there are two independent solutions.

\section{Boundary Value Problems}\label{sec:BVP}

We next discuss the solution of boundary value problems.  We solve the equation
\Beq\label{eqn:BVP}
L.X = R,
\Eeq
subject to the boundary conditions
\Beq
B.X|_{r=1} = E,
\Eeq
where $L$ and $B$ are linear operators, $R$ is a state vector, and $E$ is a state vector restricted to $r=1$.

In section~\ref{sec:dynamo}, we  initialize the magnetic vector potential from a specified magnetic field.  This is a boundary value problem, and we  use it as an example.  Specifically, we are solving
\Beq
\vec{\nabla}\vec{\times}\vec{A} & = & \vec{B}_0, \\
\vec{\nabla}\vec{\cdot}\vec{A} & = & 0.
\Eeq
Thus, our state vector is
\Beq
X \ = \ \left[
\begin{array}{c}
 A^- \\
 A^0 \\
 A^+
\end{array}
\right].
\Eeq
If we set $L$ equal to the curl operator, then we would have
\Beq
L \ &=& \ \left[
\begin{array}{ccc}
 0 & -i\xi_\ell^+ D_{0,\ell}^- & 0 \\
 -i\xi_{\ell}^{+} D_{0,\ell-1}^+ & 0 & i\xi_{\ell}^{-}D_{0,\ell+1}^- \\
 0 & i\xi_{\ell}^-D_{0,\ell}^+ & 0
\end{array}
\right].
\Eeq
However, this gives two redundant equations for $A^0$, and cannot uniquely determine $A^\pm$ because we have not set the gauge.  Thus, we replace the third row with the gauge condition $\vec{\nabla}\vec{\cdot}\vec{A}=0$,
\Beq
L' \ &=& \ \left[
\begin{array}{ccc}
 0 & -i\xi_\ell^+ D_{0,\ell}^- & 0 \\
 -i\xi_{\ell}^{+} D_{0,\ell-1}^+ & 0 & i\xi_{\ell}^{-}D_{0,\ell+1}^- \\
 \xi_{\ell}^- D_{0,\ell-1}^+ & 0 & \xi_{\ell}^+ D_{0,\ell+1}^-
\end{array}
\right].
\Eeq

The right hand side vector is given by $\vec{B}_0$, but we need to multiply by a conversion matrix $C$ to increase the $\alpha$ index to 1, since $L$ and $L'$ are both $\alpha=1$ (all terms carry a derivative operator):
\Beq
R \ = \ \left[
\begin{array}{c}
 C_{0,\ell-1}B_0^- \\
 C_{0,\ell}B_0^0 \\
 0
\end{array}
\right].
\Eeq
Finally, we must apply boundary conditions.  We use potential boundary conditions, which we impose in the last rows of the three components of the $L'$ matrix ($\alpha_{BC}=2$).

The dynamo problem in section~\ref{sec:dynamo} starts with an initial magnetic field
\Beq
\vec{B}_0 &=& -\frac{3}{2}r \left(-1+4r^2-6r^4+3r^6\right)\left(\cos(\phi)+\sin(\phi)\right)\vec{e}_\theta \nonumber \\
&& -\frac{3}{4} r\left(-1+r^2\right)\cos(\theta)\left[3 r\left(2-5r^2+4r^4\right) \sin(\theta)  \right. \nonumber \\
&&\quad\quad\quad\quad\quad\left.+ 2\left(1-3r^2+3r^4\right)\left(\cos(\phi)-\sin(\phi)\right)\right]\vec{e}_\phi.
\Eeq

To solve the boundary value problem numerically, we represent $\vec{B}_0$ with $N_{\rm max}=31$, $L_{\rm max}=31$, $\mathcal{R}_{\rm max}=2$, and no dealiasing. We invert the $L'$ matrix for each $\ell$ using \verb!scipy!'s sparse solver (\verb!sparse.linalg.spsolve!).

For this simple problem, we can solve the problem analytically and compare to the numerical solution. The cleanest way to write $\vec{A}$ in the Coulomb gauge is in terms of a poloidal function,
\Beq
\vec{A}_{\rm analytic} \ =\ \curl \curl \left[ r\, \mathcal{P}\, \vec{e}_{r}  \right],
\Eeq
where
\Beq
\mathcal{P} &=&  P_1(r) \sin (\theta )  (\sin (\phi
   )-\cos (\phi )) +  P_2(r) (3 \cos ^{2}(\theta )-1),   
\Eeq
and
\Beq
P_{1}(r) &=&  \frac{r}{16} \left(1 - \frac{12 r^2}{5} + \frac{24 r^4}{7} - \frac{8 r^6}{3} + \frac{9 r^8}{11} \right), \\
P_{2}(r) &=& \frac{3r^{2}}{160}\left(1 - \frac{20 r^2}{7} + \frac{35 r^4}{9} - \frac{30 r^6}{11} + \frac{10 r^8}{13} \right).
\Eeq
One can check that $\vec{\nabla}\vec{\times}\vec{A}_{\rm analytic}=\vec{B}_0$ and that $\vec{A}_{\rm analytic}$ satisfies the potential boundary conditions (\eq{eqn:potential BC}).

To validate our numerical solution, we calculate the error
\Beq
\frac{\max(|A_i - A_{i,{\rm analytic}}|)}{\max |A_{i,{\rm analytic}}|},
\Eeq
where $i=r,\theta,\phi$ and the maximum is across all grid points.  We find the error is $2.2\times 10^{-14}$, $2.7\times 10^{-15}$, and $2.9\times 10^{-15}$ for the three components $r$, $\theta$, and $\phi$.  Thus we have an accurate solution to this boundary value problem.

\section{Initial Value Problems}\label{sec:IVP}

We now discuss the three benchmark problems of M14.  The three problems are posed as initial value problems in the form of \eq{eqn:IVP}.  As the code described in this paper is an extension of the Dedalus code, we  refer to it with D.  We compare our results to the Marti \& Jackson code \citep{Marti2016}, and the Hollerbach code \citep{Hollerbach2000,Hollerbach2013}. The Marti \& Jackson code (hereafter MJ) decomposes all variables into scalar functions, and then expands the scalar functions in scalar spherical harmonics in the angular directions, and in Jacobi polynomials weighted by $r^\ell$ in the radial direction.  The Hollerbach code (hereafter H) also decomposes all variables into scalar functions, and uses scalar spherical harmonics in the angular directions, but uses Chebyshev polynomials in the radial direction.  Thus, H does not explicitly impose the regularity conditions at $r=0$, unlike MJ and D.

\subsection{Comparisons to Other Codes}

\subsubsection{Timestepping}

We timestep equations of the form
\Beq \label{eq:IMEX}
M.\partial_tX + L.X = F(X).
\Eeq
We use multistep implicit-explicit (IMEX) methods.   Terms on the left hand side of equation~(\ref{eq:IMEX}) are treated by linearly-implicit methods and must be linear in the evolution variables $X$, while terms on the right hand side ($F(X)$) are treated explicitly and can include both linear or nonlinear terms.  For a general multistep IMEX integrator, the new statevector at time $n+1$ is related to the statevector at earlier times by
\Beq
\left(a_{-1}M + b_{-1}L\right).X^{n+1} = \sum_{i=0}^{N} c_i F(X^{n-i}) - a_i M . X^{n-i} - b_i L.X^{n-i}.
\Eeq
We use two different time-stepping schemes: the second-order, two-step Crank-Nicolson--Adams-Bashforth scheme, CNAB2; and the fourth order, four-step semi-implicit backwards differencing formula scheme, SBDF4 \citep[both described in][]{Wang2008}. 
 Although M14 does not precisely state what time stepper MJ or H use, they likely used the second-order Runge-Kutta scheme described in \citep{Hollerbach2000} or \citep[][although it is unclear how many iterations were used]{Marti2016}.  We  refer to this time-stepping scheme as RK2.

Two of the benchmark solutions are not stationary, and we find that the choice of timestepper and timestep size plays an important role in resolving the solutions. There is no discussion of how the timestep size is chosen in M14. Although we have run simulations with adaptive timestepping, to simplify our results and enhance reproducibility, we only report simulations with constant timesteps.

\subsubsection{Resolution and Degrees of Freedom}

It is not trivial to compare the resolutions in D simulations to resolutions in MJ or H simulations. This is because we use a triangular truncation in the radial direction, unlike MJ or H. We report our radial resolution in terms of $N_{\rm max}$. However, the number of radial modes averaged over $\ell$ is roughly
\Beq
\frac{2(N_{\rm max} + 1) - \frac{1}{2}\left(L_{\rm max}+1\right)}{2}.
\Eeq
For instance, if $N_{\rm max}+1 = \frac{1}{2}(L_{\rm max}+1)$, then the average number of radial modes is about $\frac{1}{2}(N_{\rm max}+1)$. This is complicated slightly by the regularity dependence of radial modes. Also, large $\ell$'s are associated with fewer $m$ modes than small $\ell$'s.

In contrast, MJ do not appear to use a triangular truncation, and instead appear to use a constant number of radial modes, $N_r$, for every $\ell$.  This means that the maximal radial order depends on $\ell$. If $N_r=\frac{1}{2}(L_{\rm max}+1)$, as is often the case in M14, then the highest $\ell$ mode is a polynomial with order $3N_r$. Thus, one would require $\approx 3 N_r$ grid points to prevent aliasing errors, rather than the $\approx 3 N_r/2$ grid points required to dealias when using the triangular truncation.

To make a fair comparison between the codes, we  report two quantities related to the number of radial modes. First, we  report $N_{\rm max}$, which is half the highest radial order. This is analogous to reporting the angular resolution in terms of $L_{\rm max}$. For H, $N_{\rm max}$ is equal to the number of radial modes.  We  also report the total number of spatial degrees of freedom, or DoF.

\subsubsection{Energy Calculations}

A main output of the benchmark problems are the energies of the equilibrated states. However, one must take care to accurately calculate volume integrals of quantities like the energy, lest error in the volume integral itself dominate the reported results. The weights of the scalar spherical harmonics is $\sin(\theta)$, and since we use their quadrature nodes for the $\theta$ grid, we can use their quadrature weights to calculate angular integrals with spectral accuracy. Similarly, the weights of the $Q$ polynomials is $r^2$, and the quadrature weights can be used again to calculate radial integrals with spectral accuracy. Explicitly, we calculate the kinetic energy with
\Beq\label{eqn:energy}
KE = \frac{1}{2}\sum_{\vec{r}} w_\phi w_r w_\theta |\vec{u}(\vec{r})|^2,
\Eeq
where the sum is over each point on the $\vec{r}=(\phi,\theta,r)$ grid. The quadrature weights are $w_\phi=2\pi/N_\phi$, and $w_r$, $w_\theta$ derived from their respective Gaussian quadrature.

In contrast, the polynomials used in H \& MJ have an integral weight of $(1-r^2)^{-1/2}$.  Thus, to calculate integrals via quadrature, one must also include a factor of $r^2\sqrt{1-r^2}$ in the sum in \eq{eqn:energy}. However, $\sqrt{1-r^2}$ is not analytic at $r=1$, so this reduces the accuracy of the integration scheme to second order. It is possible to have very accurate solutions, but to report inaccurate energies due to a low order integration scheme. In the hydrodynamics benchmark problem, H \& D converge to the solutions with the same energy at fairly low resolution. This suggests H is not using this quadrature scheme to calculate the energy. However, MJ converges much more slowly. We hypothesize this is not due to inaccuracies in their solution, but instead due to inaccuracies in their integration scheme used for measuring $KE$.

\subsection{Hydrodynamics Problem}\label{sec:hydro}

The simplest problem in M14 solves the incompressible hydrodynamics problem with imposed velocity boundary conditions (benchmark 3),
\Beq
\partial_t\vec{u}+\vec{\nabla}p - \nu\nabla^2\vec{u} &=& -\vec{u}\vec{\cdot}\vec{\nabla}\vec{u} - 2 \Omega \,\vec{e}_z\vec{\times}\vec{u}, \\
\vec{\nabla}\vec{\cdot}\vec{u} &=& 0.
\Eeq
The terms on the left of the equals sign are timestepped implicitly, whereas the terms on the right of the equals sign are timestepped explicitly.  The boundary conditions at $r=1$ are $\vec{u}=\vec{u}_0$, where
\Beq
u_{0,\theta} &=&-u_0\cos(\theta)\cos(\phi), \\
u_{0,\phi} & = & u_0\sin(\phi).
\Eeq
Following M14, we take $\nu=10^{-2}$, $\Omega=10$, and $u_0=\sqrt{3/(2\pi)}$.

We evolve the variables $u^-$, $u^0$, $u^+$, and $p$, where $\vec{u}$ is written in terms of the three regularity classes.  We report the matrices used for the problem in Appendix~\ref{sec:matrices_hydro}.

\begin{figure}
  \centerline{\includegraphics[width=3.4in]{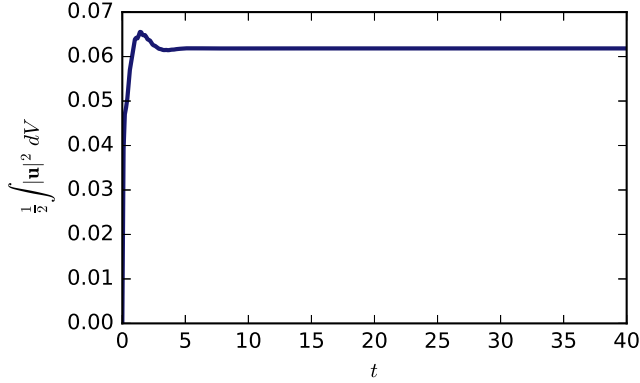}}
  \caption{The energy as a function of time in the incompressible hydrodynamics problem.  The spatial resolution is $N_{\rm max}=L_{\rm max}=31$ with $\alpha_{BC}=2$.  For timestepping we use CNAB2 with a timestep of $10^{-2}$.  The energy asymptotes to $\textit{KE}_h=0.06183074756$.}
\label{fig:hydro_E}
\end{figure}

The simulation is initialized with zero initial flow, but quickly reaches a stationary equilibrium state.  To quantitatively describe this state, we calculate the volume-integrated kinetic energy,
\Beq
\textit{KE} = \frac{1}{2} \int  |\vec{u}|^2 \ dV
\Eeq
Figure~\ref{fig:hydro_E} shows the kinetic energy as a function of time.  We run simulations to $t=40$ so we can minimize any transient effects from our initial condition.  The simulations are run with $N_{\rm max}=L_{\rm max}$ and $\mathcal{R}_{\rm max}=3$.  The low-resolution simulations were dealiased, but the higher resolution simulations were not because they are already well-resolved.  We ran with both $\alpha_{BC}=0$ and $\alpha_{BC}=2$. High resolution simulations gave identical energies for both values of $\alpha_{BC}$.  For timestepping we use a CNAB2 with a constant timestep $\Delta t$ of $2\times 10^{-2}$ or $10^{-2}$.

%\begin{centering}
\begin{table}
\centering
\begin{tabular}{cccrccrl}\hline
code & $N_{\rm max}$ & $L_{\rm max}$ & DoF & $\alpha_{BC}$ & DA & $\Delta t$ & $\textit{KE}$  \\ \hline \hline
D & 11 & 11 & 785 & 0 & Y & $0.02$ & \underline{0.06183}488623 \\
D & 15 & 15 & 1\,732 & 0 & Y & $0.02$ &   \underline{0.0618307}5192 \\
D & 23 & 23 & 5\,422 & 0 & Y & $0.02$  &  \underline{0.06183074756} \\
%D & 31 & 31 & 12\,360 & 0 & Y & $0.02$ & \underline{0.06183074756} \\
D & 31 & 31 & 12\,360 & 2 & N & $0.02$ & \underline{0.06183074756} \\
D & 31 & 31 & 12\,360 & 2 & N & $0.01$ & \underline{0.06183074756} \\
D & 63 & 63 & 93\,072 & 2 & N & $0.02$ & \underline{0.06183074756} \\ \hline
H & 12 & 11$^*$ & 600 & N/A & Y & ? & \underline{0.06183}2 \\
H & 15 & 15$^*$ & 1\,215 & N/A & Y & ? &  \underline{0.06183}1 \\
H & 24 & 23$^*$ & 4\,320 & N/A & Y & ? & \underline{0.06183}1 \\ \hline
MJ & 23 & 23 & 3\,600 & -1/2 & Y & ? & \underline{0.0618}485 \\
MJ & 31 & 31 & 8\,448 & -1/2 & Y & ? & \underline{0.06183}38 \\
MJ & 63 & 63 & 64\,480 & -1/2 & Y & ? & \underline{0.0618}286 \\
\end{tabular}
\caption{Kinetic energy resolution test for the hydrodynamics test problem in M14. The correct digits for each solution are underlined. The column DA indicates whether or not the simulation was dealiased.  We find spatial and temporal convergence to ten decimal places.  We also report the kinetic energy for several MJ and H simulations reported in M14.  No timestep size is reported for those simulations. $^*$For $L_{\rm max}=11$, $15$, and $23$, H uses $M_{\rm max}=4$, $5$, and $8$ respectively.} \label{tab:hydro}
\end{table}
%\end{centering}

Even at late times, there are small changes in the kinetic energy.  However, the kinetic energy is constant to ten decimal places between $t=35$ and $t=40$, so we report the values to ten decimal places.  The results are reported in table~\ref{tab:hydro}.  We find temporal and spatial convergence to ten decimal places, and find $\textit{KE}_h=0.06183074756$. We can reach this converged solution at a resolution of $N_{\rm max}=L_{\rm max}=23$ and timestep size of $\Delta t=0.02$. At high resolutions, we do not find any differences in the kinetic energy in simulations with or without dealiasing, or with $\alpha_{BC}=0$ or $\alpha_{BC}=2$.  At low resolutions, we find that the energies were closer to $\textit{KE}_h$ when we used $\alpha_{BC}=0$ than when we used $\alpha_{BC}=2$.  The algorithm of MJ corresponds to $\alpha_{BC}=-1/2$, whereas there is no equivalent parameter for H.

We also report the kinetic energy of several MJ and H simulations reported in M14.  They do not report their timestep size.  Our simulations are consistent with the values reported by H. With similar numbers of degrees of freedom, H appears to be slightly more accurate than our D simulations. On the other hand, our simulations appear to be significantly more accurate than those of MJ.  For this problem, we find no difference in the energy of the stationary state for simulations with different timesteps.  This is not the case for the next two problems.

\subsection{Convection Problem}\label{sec:conv}

Next we consider the rotating convection problem of M14 (benchmark 1).  For this problem, the equations are
\Beq
E \left(\partial_t - \nabla^2\right)\vec{u} +\vec{\nabla}p & = & -E\,\vec{u}\vec{\cdot}\vec{\nabla}\vec{u} + \textit{Ra}\, T \,  \vec{r} - \vec{e}_z\vec{\times}\vec{u}, \\
\vec{\nabla}\vec{\cdot}\vec{u} & = & 0, \\
\left( \textit{Pr}\, \partial_t - \nabla^2\right) T & = & S - \textit{Pr}\, \vec{u}\vec{\cdot} \vec{\nabla}T.
\Eeq
The Ekman number $E=3\times 10^{-4}$, the Rayleigh number $\textit{Ra} = 95$, the Prandtl number $\textit{Pr}=1$, and the temperature source term $S=3$.  The temperature has an equilibrium base state of $T=0.5(1-r^2)$. The vector $  \vec{r} \equiv r \vec{e}_{r}$ represents the full radial vector, and the gravity is linearly increasing, as is appropriate to a self-gravitating incompressible sphere like the Earth's core. To reach the appropriate solution, we initialize the problem with the temperature initial condition specified in M14:
\Beq
T = \frac{1}{2}\left(1 - r^2\right) + \frac{1}{8}\times10^{-5}\sqrt{\frac{35}{\pi}}r^3\left(1-r^2\right)\left(\cos(3\phi) + \sin(3\phi)\right)\sin^3(\theta).
\Eeq
The initial velocity is taken to be zero.  The boundary conditions are impenetrable and stress-free for the velocity, and fixed temperature.

We evolve the variables $u^-$, $u^0$, $u^+$, $p$, and $T$. As for the hydrodynamics benchmark, $\vec{u}$ is written in terms of the three regularity classes.  We report the matrices used for this problem in Appendix~\ref{sec:conv matrices}.

\begin{figure}
  \centerline{
\includegraphics[width=2in]{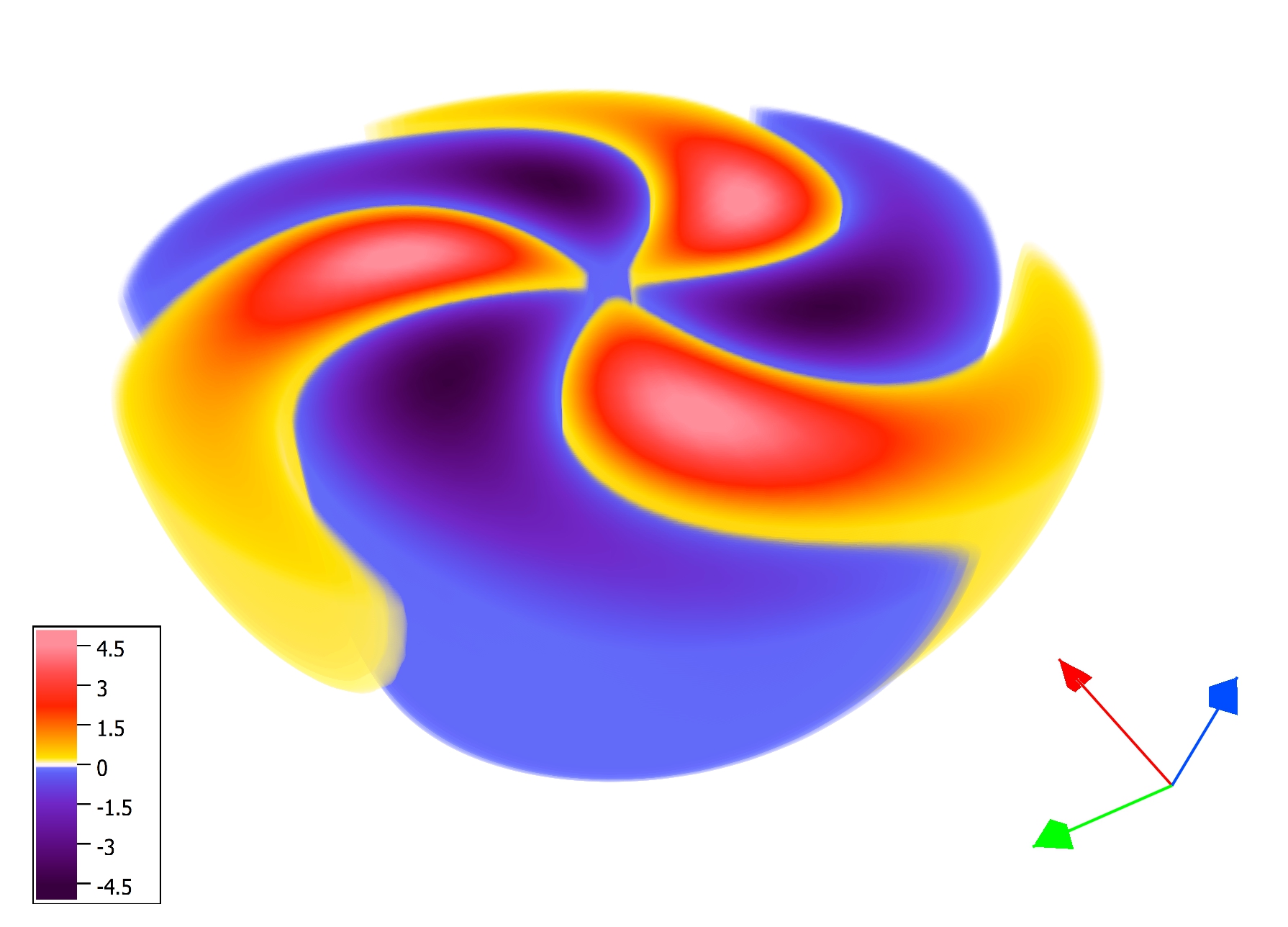}
\includegraphics[width=2in]{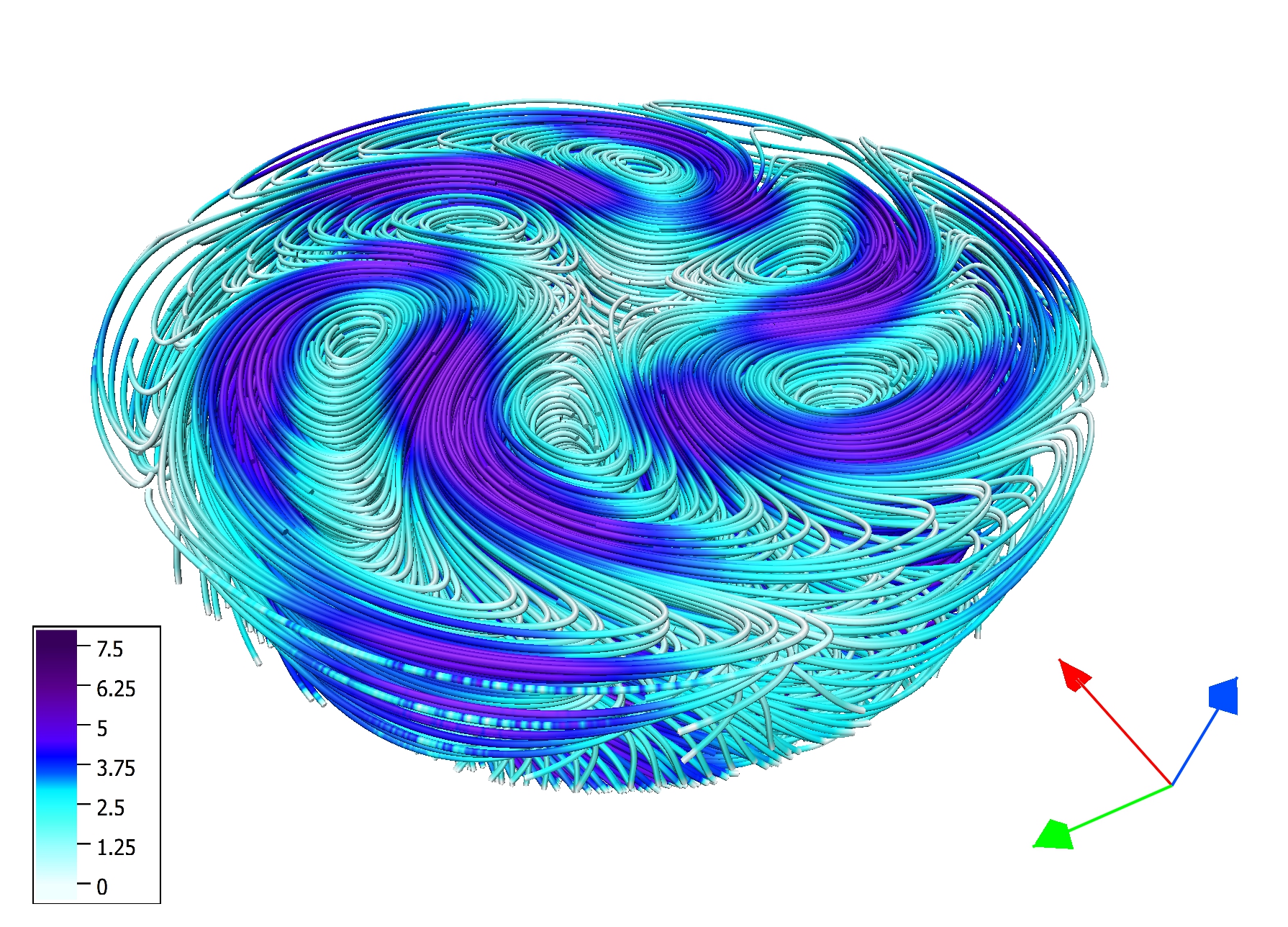}}
  \caption{Volume renderings of flows in rotating convection problem, showing characteristic $m=3$ travelling wave pattern.  Radial velocity $u_r$ is shown at left, while streamlines of the 3-D flow are shown at right, colored by the magnitude of velocity.  In both renderings, the upper half of the sphere has been cut away, showing an equatorial slice with columns descending below to the south pole.  The blue arrow is aligned with the rotation axis, pointing north.  Both images are created from the same vantage point.  Volume and streamline renderings created using Vapor \citep{vapor1, vapor2}.}
\label{marti conv flow}
\end{figure}

With this initial condition, the fluid is expected to evolve to a traveling wave state with constant kinetic energy.  As in the hydrodynamics problem, we  report the volume-integrated kinetic energy $\textit{KE}$ in Table~\ref{tab:conv}.   The simulations are run with $N_{\rm max}=L_{\rm max}$ and $\mathcal{R}_{\rm max}=3$. The low-resolution simulations were dealiased, but the higher resolution simulations were not because they are already well-resolved. We ran with both $\alpha_{BC}=0$ and $\alpha_{BC}=2$. For timestepping, we use either CNAB2 or SBDF4 with constant timesteps.  We run all simulations for 20 diffusion times.
 
The structure of the equilibrated-flow, shown in Figure~\ref{marti conv flow}, is an $m=3$ travelling wave.  The structure and amplitude of the flow match those shown in M14.  The radial $u_r$ and azimuthal $u_\phi$ flow are symmetric across the equator while the latitudinal $u_\theta$ flow is antisymmetric about the equator; as such we show a half-hemisphere, containing the equator and the south pole, in the volume and streamline renderings.  The spiralling nature of the flow is visible in the streamline rendering of Figure~\ref{marti conv flow}, with the flow dominated by $u_r$ and $u_\phi$ and with slower flows along the rotation axis.    This snapshot is taken from the equilibrated state at $t=20$ of our D simulation with $L_{\rm max}=31$, $N_{\rm max}=31$, $\alpha_{BC}=2$, and $\Delta t = 10^{-5}$ with the SBDF4 timestepper.  

\begin{figure}
  \centerline{\includegraphics[width=7.1in]{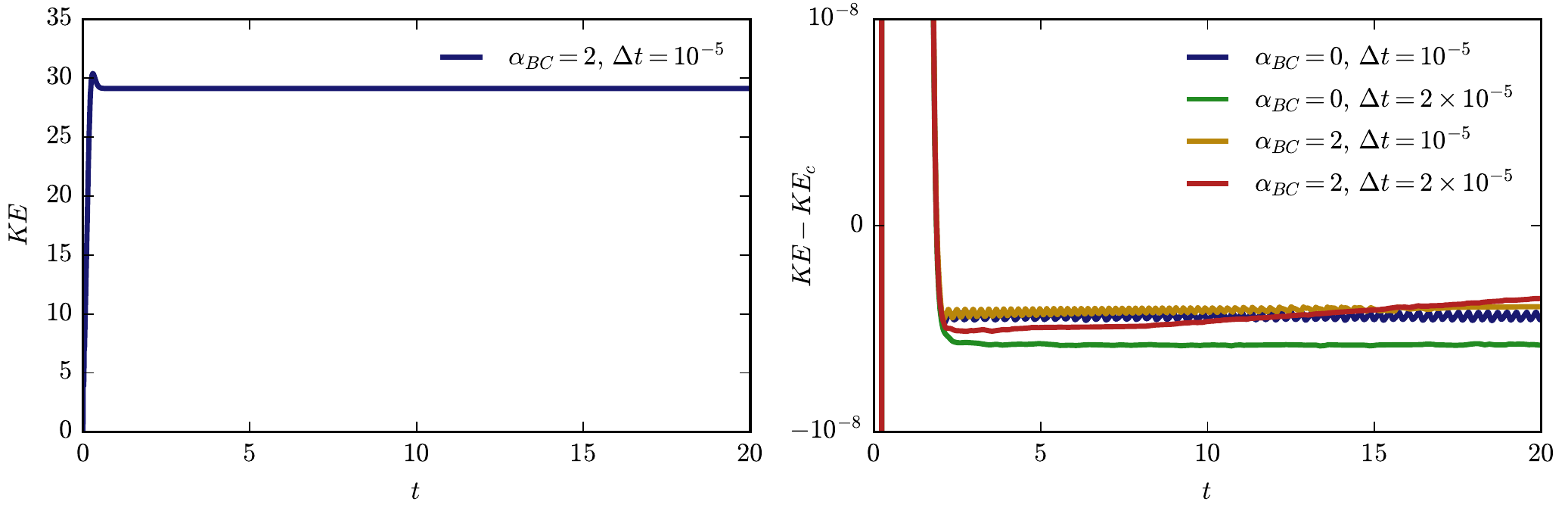}}
  \caption{The energy as a function of time in the rotating convection problem.  All simulations have resolution of $N_{\rm max}=L_{\rm max}=31$ and use the SBDF4 timestepper.  The left panel shows the full time evolution for a simulation with $\alpha_{BC}=2$ and timestep size of $10^{-5}$.  The final kinetic energy (to ten decimal places) is $\textit{KE}_c=29.12045489$.  The right panel shows secular variations in simulation with different timestep size and $\alpha_{BC}$.}
\label{fig:conv_E}
\end{figure}

Figure~\ref{fig:conv_E} shows the kinetic energy as a function of time.  The left panel shows that after a few diffusion times, the kinetic energy becomes approximately constant, indicating that we have reached the traveling wave solution.  However, the right panel shows that there is different secular behavior for different simulations.  In our simulations with larger timestep size ($\Delta t\geq 2\times 10^{-5}$) and $\alpha_{BC}=2$, we find a secular energy growth of $\sim 10^{-9}$ over the course of the simulation, similar to the red curve.  The simulations with smaller timestep size or $\alpha_{BC}=0$ show regular (blue curve) or irregular (green curve) low amplitude oscillations.  Because of these oscillations and secular variation, we report the kinetic energy at $t=20$ to ten decimal places.

%\begin{centering}
\begin{table}
\begin{tabular}{cccrcccrl}\hline
code & $N_{\rm max}$ & $L_{\rm max}$ & DoF & $\alpha_{BC}$ & DA & TS & $\Delta t$ & $\textit{KE}$  \\ \hline \hline
D & 15 & 15 & 1\,732 & 0 & Y & SBDF4 & $8\times10^{-5}$ & \underline{29.1}3102161 \\
D & 23 & 23 & 5\,422 & 0 & Y & SBDF4 & $8\times10^{-5}$ & \underline{29.12045}664 \\
D & 31 & 31 & 12\,360 & 2 & N & SBDF4 & $8\times 10^{-5}$ & \underline{29.120454}48 \\
D & 31 & 31 & 12\,360 & 2 & N & SBDF4 & $4\times 10^{-5}$ & \underline{29.1204548}6 \\
D & 31 & 31 & 12\,360 & 2 & N & SBDF4 & $2\times 10^{-5}$ & \underline{29.12045489} \\
%D & 31 & 31 &12\,360 & 0 & Y & SBDF4 & $2\times 10^{-5}$ & \underline{29.1204548}8 \\
D & 31 & 31 & 12\,360 & 2 & N & SBDF4 & $10^{-5}$ & \underline{29.12045489} \\
%D & 31 & 31 &12\,360 & 0 & Y & SBDF4 & $10^{-5}$ & \underline{29.12045489} \\
%D & 31 & 31 & 12\,360 & 2 & N & SBDF4 & $5\times 10^{-6}$ & \underline{29.12045489} \\
D & 63 & 63 & 93\,072 & 2 & N & SBDF4 & $10^{-5}$ & \underline{29.12045489} \\
%D & 63 & 63 & 93\,072 & 2 & N & SBDF4 & $5\times 10^{-6}$ & \underline{29.12045489} \\
D & 31 & 31 & 12\,360 & 2 & N & CNAB2 & $8\times 10^{-5}$ & \underline{29.12}578006 \\
D & 31 & 31 & 12\,360 & 2 & N & CNAB2 & $4\times 10^{-5}$ & \underline{29.12}178362 \\
D & 31 & 31 & 12\,360 & 2 & N & CNAB2 & $2\times 10^{-5}$ & \underline{29.120}78675 \\
D & 31 & 31 & 12\,360 & 2 & N & CNAB2    & $10^{-5}$ & \underline{29.120}53781 \\ \hline
%D & 31 & 31 & 12\,360 & 2 & N & MCNAB2 & $10^{-5}$ & \underline{29.120}54869 \\ \hline
H & 12 & 23 & 3\,600 & N/A & Y & RK2 & ? & \underline{29.1}1784 \\
H & 16 & 31 & 8\,448 & N/A & Y & RK2 & ? & \underline{29.120}54 \\
H & 31 & 63 & 64\,480 & N/A & Y & RK2 & ? & \underline{29.120}53 \\ \hline
MJ & 16 & 15 & 1\,088 & -1/2 & Y & RK2 & ? & \underline{29}.08502 \\
MJ & 24 & 23 & 3\,600 & -1/2 & Y & RK2 & ? & \underline{29.12}178 \\
MJ & 32 & 31 & 8\,448 & -1/2 & Y & RK2 & ? & \underline{29.120}64 \\
MJ & 63 & 63 & 64\,480 & -1/2 & Y & RK2 & ? & \underline{29.120}68
\end{tabular}
\caption{Kinetic energy at $t=20$ for the rotating convection test problem in M14.  The correct digits for each solution are underlined. The column TS lists the timestepper used for each simulation.  We find spatial and temporal convergence to ten decimal places.  We also report the kinetic energy for several H and MJ simulations reported in M14.  No timestep size was reported for those simulations.} \label{tab:conv}
\end{table}
%\end{centering}

Although the kinetic energy of the traveling wave is close to constant, the simulation must resolve the advection of the wave around the domain.  We find that the choice of timestepper and timestep size plays an important role in determining the kinetic energy of the traveling wave state.  In table~\ref{tab:conv} we report the kinetic energy of the traveling wave with different simulation parameters.  We achieve spatial and temporal convergence to ten decimal places, and find the kinetic energy to be $\textit{KE}_c=29.12045489$.  We can reach this level of accuracy with the fourth order timestepper SBDF4 with timestep size of $2\times 10^{-5}$ and a spatial resolution of $N_{\rm max}=L_{\rm max}=31$ with $\alpha_{BC}=2$.  At low resolution, we find that simulations with $\alpha_{BC}=0$ reach convective states with energies closer to $KE_c$ than simulations with $\alpha_{BC}=2$.

However, our value of the kinetic energy is inconsistent with the H and MJ values reported in M14.  We believe the discrepancy is due to timestepping. Both use a second order scheme, RK2, which is less accurate than SBDF4. To test this, we ran simulations with the second order timestepper, CNAB2.

\begin{figure}
  \centerline{\includegraphics[width=3.4in]{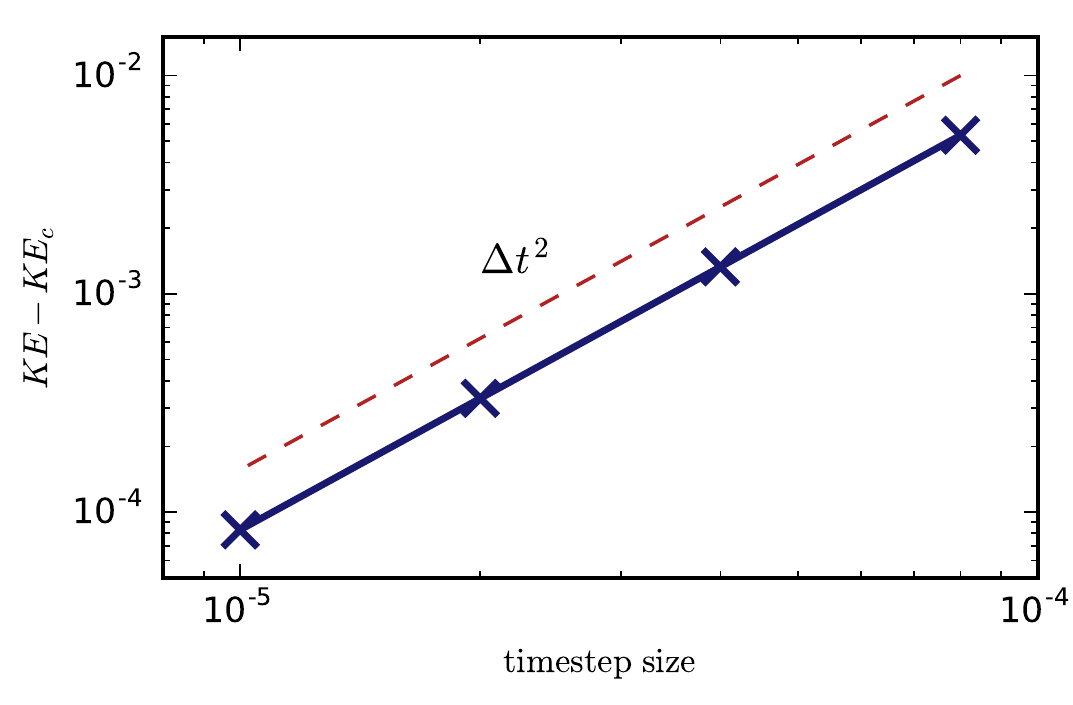}}
  \caption{The error in the kinetic energy, $\textit{KE}-\textit{KE}_c$ as a function of timestep size with the second order CNAB2 timestepper with $N_{\rm max}=L_{\rm max}=31$.  The dashed line shows a $\Delta t^2$ curve, which follows the same trend as the error.}
\label{fig:conv_E_ts}
\end{figure}

Simulations with the second order timestepper do not show temporal convergence (to ten decimal places) with timestep sizes greater than or equal to $10^{-5}$.  We find that the kinetic energies are converging to $\textit{KE}_c$ like $\Delta t^2$ (figure~\ref{fig:conv_E_ts}).  This indicates that the dominant errors in the simulations are due to timestepping.  The kinetic energies reported by Marti and Hollerbach are similar to the kinetic energies of our simulations with timestep size $10^{-5}$ and $2\times10^{-5}$.  Thus, our results are consistent with those reported by the Marti and Hollerbach codes when we use a low order timestepper. Since the CNAB2 timestepper has some well known flaws \citep{Ascher1995}, we also test the L-stable Modified CNAB2 (MCNAB2) timestepper of \citep{Wang2008} at $\Delta t=10^{-5}$. Evidently, the flaws are minor at worst.

This sensitivity of the kinetic energy to the details of timestepping make it difficult to determine the accuracy of the spatial discretization using this problem.  Temporal convergence studies, as we have done here, are necessary to distinguish spatial from temporal errors.

\subsection{Dynamo Problem}\label{sec:dynamo}

The last problem we  discuss is the rotating convective dynamo problem of M14 (benchmark 2). This is the most challenging of the three benchmark problems. We solve the equations
\Beq
\left( \textit{Ro}\,\partial_t - E \nabla^2\right)\vec{u} +\vec{\nabla}p & = & -\textit{Ro}\,\vec{u}\vec{\cdot}\vec{\nabla}\vec{u} + q \,\textit{Ra}\, T \,  \vec{r} - \vec{e}_z\vec{\times}\vec{u} + \vec{B}\vec{\cdot}\vec{\nabla}\vec{B}, \\
\vec{\nabla}\vec{\cdot}\vec{u} & = & 0, \\
\left( \partial_t - q \nabla^2\right) T & = & S - \vec{u}\vec{\cdot} \vec{\nabla}T, \\
(\partial_t-\nabla^2)\vec{A} + \vec{\nabla}\Phi & = & \vec{u}\vec{\times}\vec{B}, \\
\vec{\nabla}\vec{\cdot}\vec{A} & = & 0,
\Eeq
where $\vec{A}$ is the magnetic vector potential in the Coulomb gauge, and can be used to calculate the magnetic field $\vec{B}$ using $\vec{\nabla}\vec{\times}\vec{A}=\vec{B}$.  The scalar potential $\Phi$ enforces the Coulomb gauge constraint.  We solve for $u^-$, $u^0$, and $u^+$ (the three regularity components of $\vec{u}$), $p$ and $T$, $A^-$, $A^0$, and $A^+$ (the three regularity components of $\vec{A}$), and $\Phi$.  The exact implementation is described in Appendix~\ref{sec:dynamo matrices}.  As in M14, we set the magnetic Rossby number $\textit{Ro}=\frac{5}{7}\times10^{-4}$, the Ekman number $E=5\times10^{-4}$, the Roberts number $q=7$, the Rayleigh number $\textit{Ra}=200$, and the temperature source term to $S=3q=21$.  For boundary conditions we use impenetrable, stress-free boundary conditions for the velocity, fixed temperature, and potential boundary conditions for the magnetic field (see section~\ref{sec:diffusion}).

We use the temperature and velocity initial condition described in M14.  Because we evolve the magnetic vector potential and not the magnetic field directly, we solve a boundary value problem to initialize the vector potential (see section~\ref{sec:BVP}).  We run simulations with $N_{\rm max}=L_{\rm max}$ and $\mathcal{R}_{\rm max}=3$.  The low-resolutions simulations were dealiased, but the higher resolution simulations were not because they are already well-resolved. We ran with both $\alpha_{BC}=0$ and $\alpha_{BC}=2$, and again we found more accurate solutions at low resolutions using $\alpha_{BC}=0$. For timestepping, we use either CNAB2 or SBDF4 with constant timesteps.

\begin{figure}
 \centerline{\includegraphics[width=7.1in]{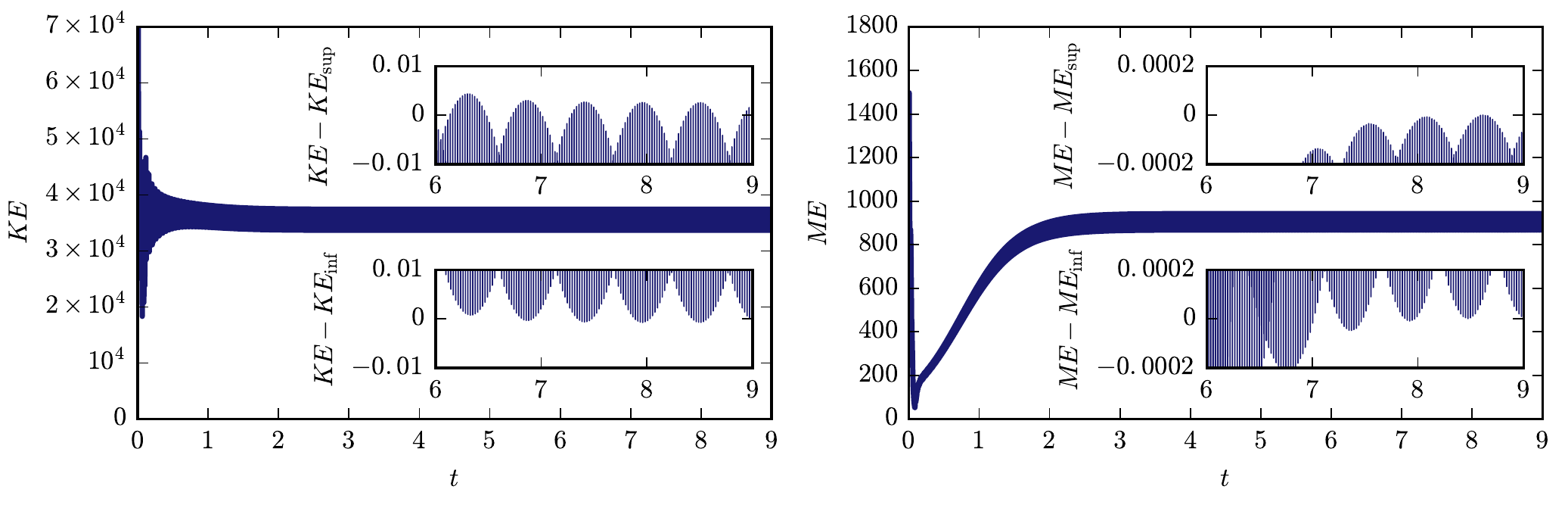}}
  \caption{The kinetic energy (left) and magnetic energy (right) of the convective dynamo benchmark with resolution $N_{\rm max}=L_{\rm max}=63$ with $\alpha_{BC}=2$.  We use the SBDF4 timestepper with timesteps of $2.5\times 10^{-6}$. The kinetic and magnetic energies oscillate rapidly, which leads to an extended opaque region on the plot.  In each inset, we zoom in onto the top or bottom of each oscillatory region, and plot the deviation of the energies from their limit suprema---they might not go to zero because we round the limit suprema.  For the kinetic energy, we use $\textit{KE}_{\rm inf} = 33681.31$ and $\textit{KE}_{\rm sup}=37444.32$ (to within an accuracy of $10^{-2}$).  For the magnetic energy, we use $\textit{ME}_{\rm inf} = 867.7413$ and $\textit{ME}_{\rm sup}=943.4111$ (to within an accuracy of $10^{-4}$).  These can be combined to calculate $\overline{\textit{KE}}$, $\Delta \textit{KE}$, etc.}
\label{fig:dynamo_E}
\end{figure}

The system approaches an oscillating dynamo solution.  The kinetic and magnetic energy,
\Beq
\textit{ME} = \frac{1}{2\textit{Ro}}\int  |\vec{B}|^2 \ dV,
\Eeq
undergo regular variations over the oscillation period.  We plot the kinetic and magnetic energy in figure~\ref{fig:dynamo_E}.  After an initial transient of $\sim 2$ magnetic diffusion times, the system approaches an oscillating dynamo solution.  The kinetic and magnetic energy oscillate rapidly in this state. For some of the highest resolution simulations, we restarted the simulation from a low-resolution simulation evolved beyond the initial transient.

\begin{figure}
  \centerline{\includegraphics[width=3in]{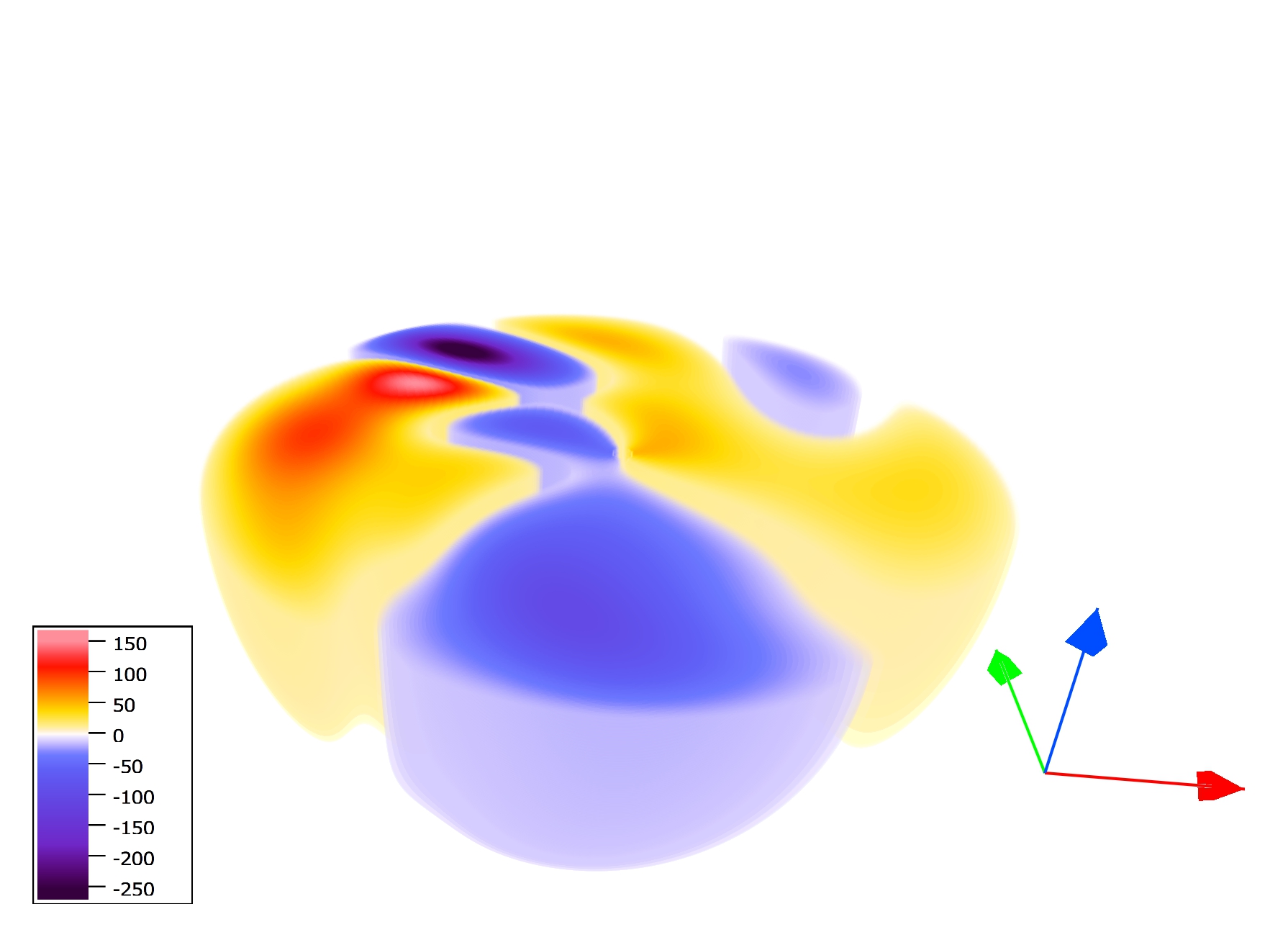}\includegraphics[width=3in]{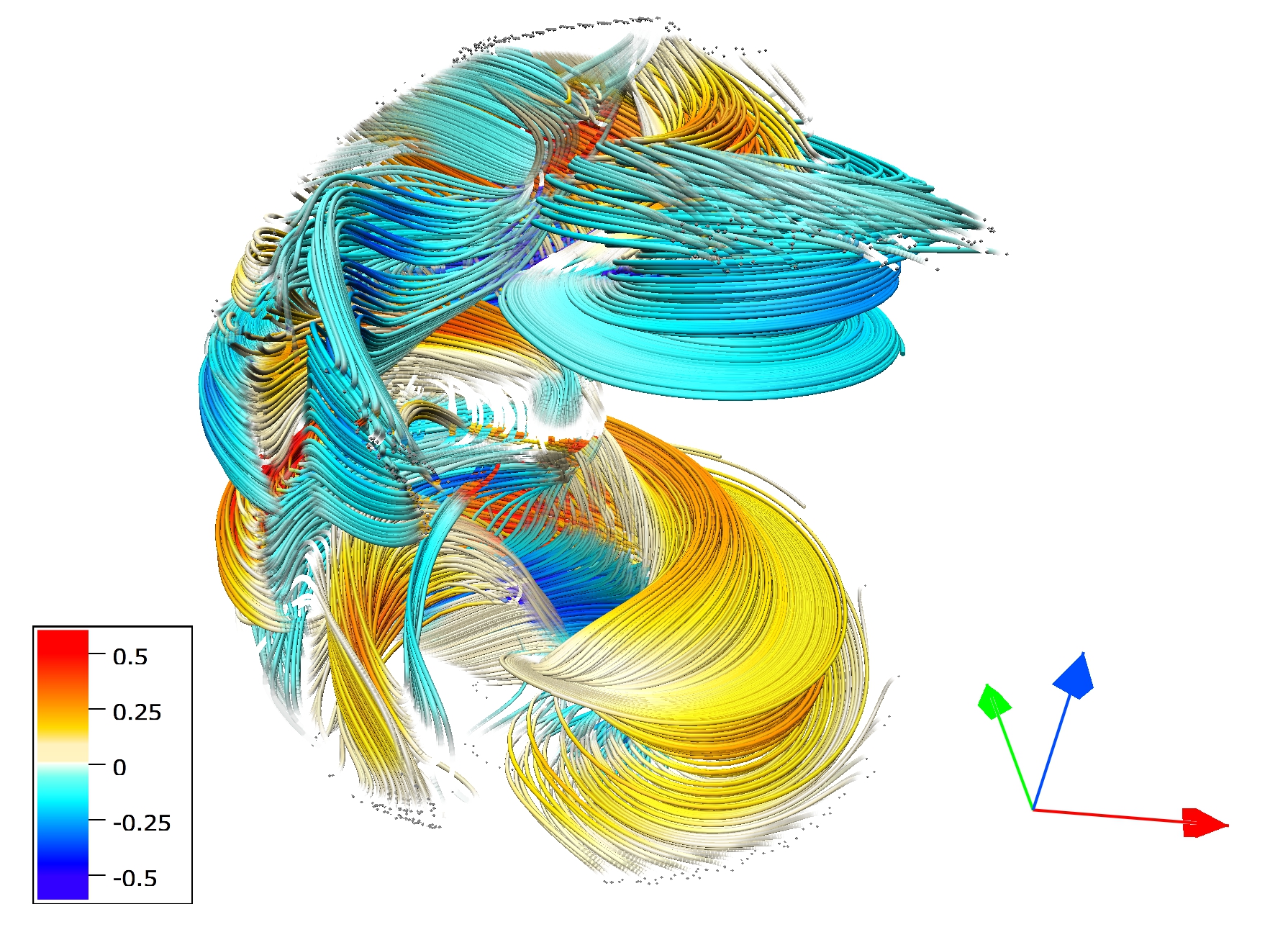}}
  \caption{Volume rendering of radial velocity (left) and magnetic field lines (right) at a late time in the dynamo problem.  In the radial velocity $u_r$ volume rendering (left), the upper half of the sphere has been cut away, showing an equatorial slice with columns descending below to the south pole and a marked asymmetry in flow structures.  The transfer function is asymmetric, but with both color ranges diverging from zero flow.  In the magnetic field line rendering (right), the full volume is shown from the same vantage point as the radial flow rendering.  Field lines are seeded in the strongest regions of $|B_\phi|$, and field lines are colored by the phi component $B_\phi$.  Here, in contrast to $u_r$, the transfer function and the field $B_\phi$ are both symmetric.   The blue arrow is aligned with the rotation axis, pointing north.  Volume and field rendering created using Vapor \citep{vapor1, vapor2} from the same vantage point.}
\label{marti dynamo flow field}
\end{figure}

A volume rendering of a late state of the oscillating dynamo simulation with $N_{\rm max}=L_{\rm max}=63$, $\alpha_{BC}=2$, no dealiasing, and timestep size of $2.5\times 10^{-6}$ using the SBDF4 scheme is shown in Figure~\ref{marti dynamo flow field}.  Shown are the radial velocity field $u_r$ in half-domain rendering (as in Figure~\ref{marti conv flow}), and magnetic field line rendering in the full domain.  In contrast to the rotating convection problem, the dynamo problem has asymmetries in the velocity field, with a region of strong up and downflow visible near the front of the rendering.  The strong fields themselves are found in the region of strong flow.  These volume and field line renderings were created using Vapor\footnote{See: \url{http://www.vapor.ucar.edu/}}. This snapshot is taken from the equilibrated oscillating state at $t=5.5$ of our high resolution D simulation with $L_{\rm max}=127$, $N_{\rm max}=63$, $\alpha_{BC}=2$, and $\Delta t = 1.25\times10^{-6}$ with the SBDF4 timestepper.  At $t=5.5$, $\textit{KE}(5.5) = 33860.4$ and $\textit{ME}(5.5)=933.404$ and the solution is the decreasing magnetic energy phase of the oscillation.

\subsubsection{Energy Diagnostics}

Because the energy is not constant, there are different ways to characterize the system.  M14 decomposes the energy into its two dominant temporal Fourier modes
\Beq\label{eqn:KE expansion}
\textit{KE}(t) &=& C_k + A_k\sin(2\pi f t +\zeta_k) + \ldots, \\
\textit{ME}(t) &=& C_m + A_m\sin(2\pi f t +\zeta_m) + \ldots. \label{eqn:ME expansion}
\Eeq
Most of the energy are in these modes, and the higher harmonics of $f$.  The $2f$ harmonic contains a few percent of the energy of the time series.  To perform the decomposition, we take each energy time series over the final half a diffusion time of the simulation.  Then we identify the first and last energy maximum in the time series. We truncate the time series so it ranges from the time of the first maximum to the last output time before the last maximum. This makes the time series approximately periodic. We then take the Fourier transform. The amplitudes of the first and second peak give $C$ and $A$.

Unfortunately, this approach is sensitive to various choices in the algorithm. For instance: using the energy time series for a full diffusion time or a half of a diffusion time; running the algorithm with a different output cadence; or, different truncation methods, all give different coefficients $C$ and $A$.  It is difficult to calculate the first two coefficients of an expansion to high precision (e.g., $10^{-7}$), when the third coefficient has size $10^{-2}$.

We also consider a new, more robust metric to characterize the system. The simulation approaches an oscillatory state as $t\rightarrow \infty$. The dynamo oscillation has well defined minima and maxima of kinetic and magnetic energy. Thus, we characterize the solution by calculating these minima and maxima. Formally these are the limit extrema as $t\rightarrow\infty$. We define
\Beq\label{eqn:KE sum}
\overline{\textit{KE}} &=& \frac{1}{2}\left[\limsup_t(\textit{KE}\,) + \liminf_t(\textit{KE}\,)\right], \\
\Delta \textit{KE} &=& \frac{1}{2}\left[\limsup_t(\textit{KE}\,) - \liminf_t(\textit{KE}\,)\right], \\
\overline{\textit{ME}} &=& \frac{1}{2}\left[\limsup_t(\textit{ME}\,) + \liminf_t(\textit{ME}\,)\right], \\
\Delta \textit{ME} &=& \frac{1}{2}\left[\limsup_t(\textit{ME}\,) - \liminf_t(\textit{ME}\,)\right].\label{eqn:ME diff}
\Eeq
We refer to the limit extrema as $\textit{KE}_{\rm inf}$, $\textit{KE}_{\rm sup}$, and similar for the magnetic energy.

To determine the limit extrema energies, we include insets in figure~\ref{fig:dynamo_E} in which we zoom in to energy scales close to the limit extrema.  Here we can see a regular pattern in the energy extrema.  This is because the, e.g., maximal energy might occur between time steps, so we must integrate for many oscillation periods before our integration reaches a time which, by coincidence, is very close to the time of an energy extremum.  The limit extrema are rounded to $10^{-2}$ for kinetic energy and $10^{-4}$ for magnetic energy (7 digits reported in both cases).  We show below that these metrics are more robust than the $(C_{k},A_{k})$ and $(C_{m},A_{m})$ decompositions described in M14.

\subsubsection{Low-resolution Solutions}\label{sec:low_res}

\begin{figure}
  \centerline{\includegraphics[width=3.4in]{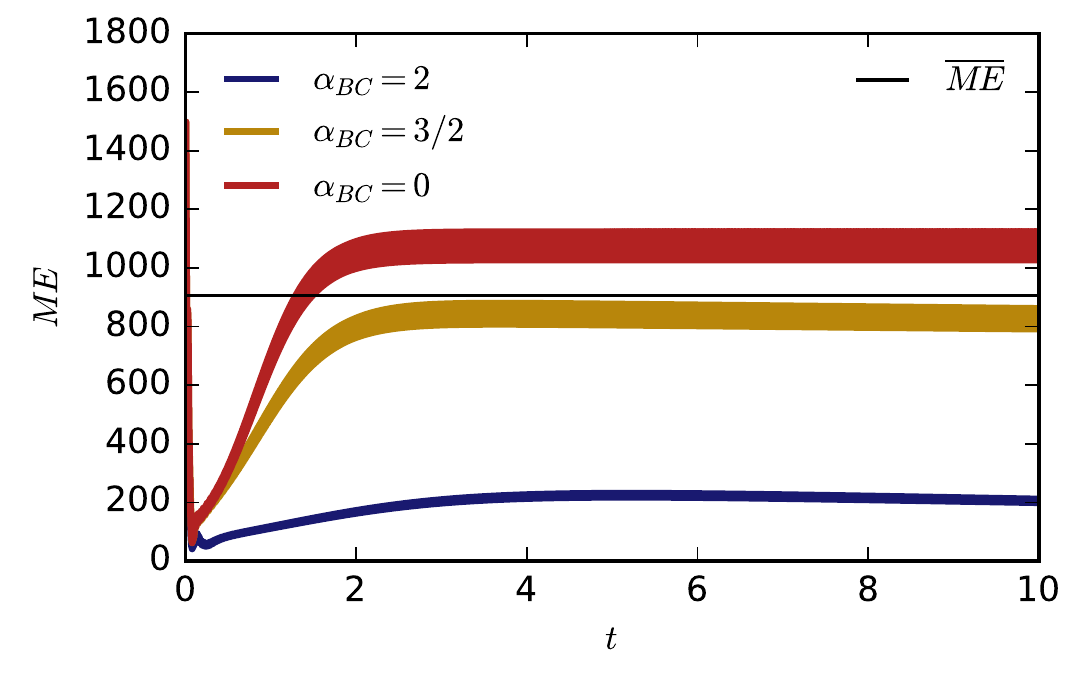}}
  \caption{The magnetic energy as a function of time for three dynamo simulations with $N_{\rm max}=L_{\rm max}=23$, but with different implementation of boundary conditions ($\alpha_{BC}=0$, $3/2$, or $2$).  Despite only very minor changes in the numerical algorithm, we find completely different solutions at these low, unresolved resolutions.  Although $\alpha_{BC}=3/2$ appears closest to the correct solution (black line), its magnetic energy is decaying secularly with time---we hypothesize its magnetic energy trends to zero as $t\rightarrow\infty$.}
\label{fig:dynamo_lres}
\end{figure}

We found the low-resolution calculations ($N_{\rm max}$ and $L_{\rm max}$ less than 63) are extremely sensitive to numerical details.  We perform a set of simulations with $N_{\rm max}=L_{\rm max}=23$, using the CNAB2 timestepper with timestep size of $5\times 10^{-6}$.  We use three different values of $\alpha_{BC}$: $0$, $3/2$, and $2$.  The magnetic energy of each solution is shown in figure~\ref{fig:dynamo_lres}.  For comparison, we also show $\overline{ME}=905.566$ (the average magnetic energy at late times) as calculated in high resolution simulations (see table~\ref{tab:dynamo}).

Although almost all simulation parameters are the same, we find completely different solutions with different values of $\alpha_{BC}$.  It may appear that the most accurate solution uses $\alpha_{BC}=3/2$.  However, a more careful inspection of the data shows that the magnetic energy decays secularly by about $2.5$ energy units every magnetic diffusion time.  The energy in the simulation with $\alpha_{BC}=2$ also decays at late times.  We suspect that in both simulations, the magnetic energy decays to zero as $t\rightarrow\infty$, i.e., they do not represent dynamo solutions.  We also find that the magnetic energy decays slowly when using $\alpha_{BC}=2$ at resolutions $N_{\rm max}=L_{\rm max}=31$ and $47$, although the decay rate becomes smaller as the resolution increases.  As shown in Figure~\ref{fig:dynamo_E}, the energy asymptotes to a constant oscillate at the medium resolution of $63$.  In contrast, the magnetic energy in the simulation with $\alpha_{BC}=0$ stays constant at late times at all resolutions we tried.

As described above, we found $\alpha_{BC}=0$ is consistently more accurate than $\alpha_{BC}=2$ at low resolutions in all our tests.  For the other problems, the difference between $\alpha_{BC}=0$ and $\alpha_{BC}=2$ was at most a few extra digits of accuracy.  However, for this problem, there is an order unity difference between the simulations with $\alpha_{BC}=0$ and $\alpha_{BC}=2$.  For this reason, we think the low resolution ($N_{\rm max}=L_{\rm max}=23$) dynamo problem is a good numerical test. It can show the limitations of a given numerical scheme and sensitivity to different methods in marginally resolved simulations. 

\subsubsection{Convergence Study}

\begin{table*}
\hspace{-0.5in}
\begin{tabular}{cccrcccrllll}\hline
code & $N_{\rm max}$ & $L_{\rm max}$ & DoF & $\alpha_{BC}$ & DA & TS & $\Delta t$ & $C_k$ & $A_k$ & $C_m$ & $A_m$  \\ \hline \hline
D & 23 & 23 & 5\,422 & 0 & Y & CNAB2 & 5e-6 & \underline{35}291.06 & \underline{18}45.57 & 1079.6746 & 47.5053 \\
D & 31 & 31 & 12\,360 & 0 & Y & CNAB2 & 5e-6 & \underline{355}41.70 & \underline{18}80.58 & \underline{9}25.0387 & \underline{3}8.5001 \\
D & 47 & 47 & 40\,076 & 0 & Y & CNAB2 & 5e-6 &  \underline{35551}.40 & \underline{18}80.13 & \underline{90}9.0677 & \underline{37.4}733 \\
D & 63 & 63 & 93\,072 & 2 & N & SBDF4 & 2.5e-6 & \underline{35551}.26 & \underline{1879}.80 & \underline{908.6}604 & \underline{37.44}34  \\
D & 127 & 63 & 226\,192 & 2 & N & SBDF4 & 2.5e-6 & \underline{35551}.16 & \underline{1879}.73 & \underline{908.6}605 & \underline{37.44}34 \\
D & 63 & 95 & 158\,680 & 2 & N & SBDF4 & 2.5e-6 & \underline{35551}.17 & \underline{1879}.80 & \underline{908.65}13 & \underline{37.44}66 \\
D & 63 & 127 & 193\,312 & 2 & N & SBDF4 & 2.5e-6 & \underline{35551}.09 & \underline{1879}.71 & \underline{908.65}22 & \underline{37.44}65 \\
D & 63 & 127 &193\,312 & 2 & N & SBDF4 & 1.25e-6 & \underline{35551}.14 & \underline{1879}.80 & \underline{908.65}15 & \underline{37.44}65 \\ \hline
H & 12 & 23 & 3\,600 & N/A & Y & RK2 & ? & \underline{35}378 & \underline{18}55 & \underline{}1043.77 & \underline{}46.16 \\
H & 16 & 31 & 8\,448 & N/A & Y & RK2 & ? & \underline{355}88 & \underline{18}85 & \underline{90}4.30 & \underline{37}.61 \\
H & 23 & 47 & 27\,048 & N/A & Y & RK2 & ? & \underline{35551} & \underline{18}80 & \underline{90}9.67 & \underline{37.4}8 \\
H & 31 & 63 & 64\,480 & N/A & Y & RK2 & ? & \underline{3555}0 & \underline{18}80 & \underline{90}9.46 & \underline{37.4}7 \\ \hline
MJ & 24 & 23 & 3\,600 & -1/2 & Y & RK2 & ? & \underline{35}141.84 & \underline{18}36.287 & \underline{}1153.695 & \underline{}51.77003 \\
MJ & 32 & 31 & 8\,448 & -1/2 & Y & RK2 & ? & \underline{355}48.95 & \underline{18}81.661 & \underline{9}22.3073 & \underline{3}8.54002 \\
MJ & 47 & 47 & 27\,048 & -1/2 & Y & RK2 & ? & \underline{35551}.33 & \underline{18}80.055 & \underline{908}.9870 & \underline{37.4}7705 \\
MJ & 63 & 63 & 64\,480 & -1/2 & Y & RK2 & ? & \underline{3555}0.93 & \underline{1879}.837 & \underline{908}.8059 & \underline{37.4}5069 
\end{tabular}
\caption{The kinetic and magnetic energy expansion coefficients (see equations~\ref{eqn:KE expansion} \& \ref{eqn:ME expansion}) for the convective dynamo test problem of M14.  The correct digits for each solution are underlined. We also include the expansion coefficients of H and MJ reported in M14. No timestep size was reported for those simulations.} \label{tab:dynamo_marti}
\end{table*}

We report the kinetic and magnetic energies in our simulations in tables~\ref{tab:dynamo_marti} \& \ref{tab:dynamo}.  Our high resolution simulations are run without dealiasing, and with a higher order timestepper.  We find no significant difference between $\alpha_{BC}=0$ and $\alpha_{BC}=2$.  Using the decomposition of the kinetic and magnetic energy into the first two Fourier modes, our simulations appear to show convergence to 5 decimal places in $C_k$, 4 decimal places in $A_k$, 5 decimal places in $C_m$ and 4 decimal places in $A_m$. Subsequent digits of each quantity vary with different spatial or temporal resolution. However, these differences are primarily due to the decomposition algorithm, rather than differences in the actual dynamo solution.

The low-resolution Dedalus simulations have similar accuracies as H and MJ.  For the lowest resolution ($N_{\rm max}=L_{\rm max}=23$), our solution is more accurate than MJ, but less accurate than H.  Of course, all three simulations are far from the correct solution.  At a resolution of $N_{\rm max}=L_{\rm max}=31$, our solution has very similar energy to MJ---H appears to be more accurate for the magnetic energy, but less accurate for the kinetic energy.  Higher resolutions cannot be easily compared to a reference solutions because the differences in energies between the different simulations are likely due to differences in the decomposition into $C$ and $A$ rather than real differences in the solutions.

%\begin{centering}
\begin{table}
\hspace{-0.5in}
\begin{tabular}{cccrcccrllrl}\hline
code & $N_{\rm max}$ & $L_{\rm max}$ & DoF & $\alpha_{BC}$ & DA & TS & $\Delta t$ & $\overline{KE}$ & $\Delta KE$ & $\overline{ME}$ & $\Delta ME$  \\ \hline \hline
D & 23 & 23 & 5\,422 & 0 & Y & CNAB2 & 5e-6 & \underline{35}301.39 & \underline{18}47.32 & 1075.5958 & 48.0285 \\
D & 31 & 31 & 12\,360 & 0 & Y & CNAB2 & 5e-6 & \underline{355}53.09 & \underline{188}2.24 & \underline{9}21.7526 & \underline{3}8.9416 \\
D & 47 & 47 & 40\,076 & 0 & Y & CNAB2 & 5e-6 &  \underline{3556}3.07 & \underline{1881}.90 & \underline{905}.9678 & \underline{37.8}694 \\
D & 63 & 63 & 93\,072 & 2 & N & SBDF4 & 2.5e-6 & \underline{35562}.82 & \underline{1881.5}0 & \underline{905.5}762 & \underline{37.83}49  \\
D & 127 & 63 & 226\,192 & 2 & N & SBDF4 & 2.5e-6 & \underline{35562}.82 & \underline{1881.5}0 & \underline{905.5}762 & \underline{37.83}49 \\
D & 63 & 95 & 158\,680 & 2 & N & SBDF4 & 2.5e-6 & \underline{35562.75} & \underline{1881.5}0 & \underline{905.565}4 & \underline{37.8392} \\
D & 63 & 127 & 193\,312 & 2 & N & SBDF4 & 2.5e-6 & \underline{35562.75} & \underline{1881}.49 & \underline{905.565}7 & \underline{37.8392} \\
D & 63 & 127 &193\,312 & 2 & N & SBDF4 & 1.25e-6 & \underline{35562.75} & \underline{1881}.49 & \underline{905.565}7 & \underline{37.8392}
\end{tabular}
\caption{The normalized sums and differences of the limit extrema of the kinetic and magnetic energy (equations~\ref{eqn:KE sum}-\ref{eqn:ME diff}) for the convective dynamo test problem of M14.  The correct digits for each solution are underlined.  We find spatial and temporal convergence to six or seven digits for each quantity.} \label{tab:dynamo}
\end{table}
%\end{centering}

Table~\ref{tab:dynamo} shows the normalized sums and differences of the limit extrema of the kinetic and magnetic energy. We can only report Dedalus simulations as this quantity was not reported in M14. Note that each quantity is different from their corresponding values in table~\ref{tab:dynamo_marti} by a few percent. This is due to the effect of higher order harmonics dropped in equations~\ref{eqn:KE expansion} \& \ref{eqn:ME expansion}.

Our fiducial simulation has resolution of $N_{\rm max}=L_{\rm max}=63$, $\alpha_{BC}=2$, is run without dealiasing, and uses the SBDF4 timestepper with constant timestep size of $2.5\times 10^{-6}$. If we increase the radial resolution to $N_{\rm max}=127$, none of the quantities change to the accuracy reported.  This indicates that the simulation is radially well-resolved to this level of accuracy with $N_{\rm max}=63$.  We then fix $N_{\rm max}=63$ and increase the angular resolution. We find the same value for $\overline{\textit{KE}}$ and $\Delta \textit{ME}$ to all digits reported between $L_{\rm max}=95$ and $L_{\rm max}=127$. However, $\Delta \textit{KE}$ differs by $10^{-2}$ and $\overline{\textit{ME}}$ differs by $3\times10^{-4}$. Finally, we checked temporal convergence by running our highest resolution simulation with time-step size of $1.25\times 10^{-6}$. This did not change the quantities to the accuracy reported. Thus, we report the converged values
\Beq\label{eqn:aveKE}
\overline{\textit{KE}} & = & 35562.75, \\
\Delta \textit{KE} & = & 1881.5, \\
\overline{\textit{ME}} &=& 905.566, \\
\Delta \textit{ME} &=& 37.8392. \label{eqn:diffME}
\Eeq
Simulations with higher angular resolution (or run with dealiasing) could increase the accuracy of $\Delta \textit{KE}$ and $\overline{\textit{ME}}$. The other quantities cannot be determined to higher accuracy without running the simulations for longer to minimize the effects of transients (see figure~\ref{fig:dynamo_E}).

\section{Conclusions}

This paper describes the implementation of a new method for the solution of a wide class of partial differential equations in a full sphere, described in Part-I. We represent tensor variables using spin-weighted spherical harmonics in the angular direction, and a scaled class of Jacobi polynomials in the radial direction. This ensures that each quantity satisfies the appropriate regularity conditions, both at the poles, and at $r=0$. We can calculate nonlinear quantities by transforming the solution from spectral space to physical space, and performing products or other operations in physical space.

To demonstrate the accuracy of this method, we first discuss a series of unit tests which test specific aspects of the code. The first is the Bessel's equation eigenvalue problem (section~\ref{sec:bessel}). This is a non-trivial problem, as Bessel functions are not polynomials. The accurate solution of Bessel's equation thus demonstrates the exponential convergence of our algorithm.

We next solve for the decaying eigenmodes of a diffusing, divergence-free vector field (section~\ref{sec:diffusion}). Here we use the tensorial nature of our algorithm to rewrite the three components of the vector into three regularity classes, each of which has different behavior as $r\rightarrow 0$. We are able to impose different boundary conditions on the eigenvalue problem: no slip or stress-free boundary conditions if the vector field is the velocity; potential, conducting, or pseudo-vacuum boundary conditions if the vector field is the magnetic vector potential associated with a magnetic field in the Coulomb gauge. In each case, we can solve the eigenvalue problem analytically in terms of Bessel functions. This allows us to precisely compare our eigenvalues to the analytical eigenvalues, and demonstrates that we correctly impose all boundary conditions.

We include an example of the solution of a boundary value problem (section~\ref{sec:BVP}). We solve for the magnetic vector potential which corresponds to a specified magnetic field. The magnetic field is a polynomial in $r$ and only includes a few spherical harmonic components in the angular direction. This problem is a useful test of our transforms. These unit tests are invaluable for code verification.

The remainder of the paper describes our solutions to three full-code, initial value problems proposed in M14. The first problem we consider is a hydrodynamics problem (M14's benchmark 3; section~\ref{sec:hydro}). The system is forced with a velocity boundary condition, which leads to a stationary, nonlinear equilibrium. Crucially, the solution is independent of the details of timestepping, which is not the case for the other two problems. This makes the problem an excellent test of the spatial discretization. It also ensures the code can correctly impose the boundary conditions, and calculate the Coriolis force and the $\vec{u}\vec{\cdot}\vec{\nabla}\vec{u}$ nonlinearity. We find excellent agreement with Hollerbach's simulations (M14). We show spatial and temporal convergence of the kinetic energy to 10 digits of precision: $\textit{KE}_h=0.06183074756$.

The second problem is a rotating convection problem (M14's benchmark 1; section~\ref{sec:conv}). The system evolves toward a traveling wave solution, whose kinetic energy is constant in time. We find that the saturated value of the kinetic energy depends on the timestepping scheme. We show that for moderate spatial resolution, the error in the kinetic energy can be dominated by timestepping errors, and decreases like $\Delta t^2$ for a second-order timestepper (figure~\ref{fig:conv_E_ts}). Thus, we find that this problem is more of a test of a code's timestepper, rather than its spatial discretization. This makes it difficult to compare to previous results in M14, as they do not provide sufficient details about their timestepping. Nevertheless, we find similar results to the Hollerbach and Marti-Jackson codes when we run with a second-order accurate timestepper. Switching to a fourth-order accurate timestepper, we are able to show spatial and temporal convergence of the kinetic energy to 10 digits of precision: $\textit{KE}_c =29.12045489$.

The last problem is a convective dynamo problem (M14's benchmark 2; section~\ref{sec:dynamo}). This is the most challenging problem in M14. This problem is very sensitive to the numerical method at low resolution, where we find that minor numerical choices can lead to completely different solutions (see figure~\ref{fig:dynamo_lres}).  

The desired end state for this problem is an oscillating dynamo solution, for which both the kinetic and magnetic energy are variable. Thus, one must decide how to characterize the solutions. In M14, the kinetic and magnetic energy time series were decomposed in a series expansion, and they report the first two terms in the series. This is not very precise, as the third term in the expansion has a relative size of $\mathcal{O}(10^{-2})$, and different methods of carrying out the decomposition may give systematically different results. Nevertheless, if we implement this decomposition, we recover solutions similar to those in M14. It is difficult to tell if differences between the codes are due to differences in the spatial discretization, the temporal discretization, or the algorithm used to preform the energy decomposition.

We also discuss a new diagnostic for this convective dynamo problem which is very precise. We calculate the limit superior and limit inferior of the kinetic and magnetic energies, and then calculate their sums and differences. This considers all possible terms in the series expansion considered in M14 and gives very precise solutions. We find temporally and spatially converged solutions to 5 or 6 digits of precision, as reported in equations~\ref{eqn:aveKE}-\ref{eqn:diffME}. We hope this new diagnostic makes this problem more useful for quantitative code comparison.

We found it difficult to compare to the solutions of the Hollerbach and Marti-Jackson code because M14 does not include some important details of the simulations, e.g., timestepping scheme or timestep size, all the details of the algorithm for calculating the volume-integrated energy, etc. This makes it unclear if different results are due to important differences between spatial discretization schemes, or simply due to different timestep sizes. To aid future comparison to the solutions we describe here, we include the source code used to run the simulations, as well as the data and analysis scripts used to generate the plots. We believe that this information makes future code comparisons much more fruitful.

We have implemented the algorithms of Part-I using aspects of the Dedalus code.  In the future, we will fully incorporate spherical geometry into the main Dedalus codebase. This will allow us to use the Dedalus equation parser to write equations out as strings, rather than manually constructing matrices and the nonlinear terms. This will also allow the user to specify complicated simulation outputs (e.g., enstrophy, Reynolds stresses, etc.) in string form. These features will make this algorithm straightforward to use for the solution of many different PDEs in spherical geometry.

\section*{Acknowledgments}

We thank Nathana\"{e}l Schaeffer for sharing some of his simulation results for the dynamo problem. DL is supported by a Hertz Foundation Fellowship, the National Science Foundation Graduate Research Fellowship under Grant No. DGE 1106400, a PCTS fellowship, and a Lyman Spitzer Jr. fellowship. GMV acknowledges support from the Australian Research Council, project number DE140101960. Computations were conducted with support by the NASA High End Computing (HEC) Program through the NASA Advanced Supercomputing (NAS) Division at Ames Research Center on Pleiades with allocations GID s1647 and s1439.

Declarations of interest: None.

\appendix

\section{Analytic Solution to Linear Diffusion Problem}\label{sec:analytic_solution}

We can solve the linear diffusion problem (equations~\ref{eqn:diffusion_vel} \& \ref{eqn:diffusion_div}, or equations~\ref{eqn:diffusion_potential} \& \ref{eqn:diffusion_gauge}) analytically by decomposing $\vec{u}$ or $\vec{A}$ into toroidal--poloidal form, e.g.,
\Beq
\label{vector-PT-solution}
\vec{u} = \vec{\nabla}\vec{\times}\left(r \mathcal{T}(r) Y_{\ell,m}(\theta,\phi)\vec{e}_r\right) + \vec{\nabla}\vec{\times}\vec{\nabla}\vec{\times}\left(r \mathcal{P}(r) Y_{\ell,m}(\theta,\phi)\vec{e}_r\right).
\Eeq
The pressure or scalar potential is decomposed such that 
\Beq
p \ = \ \varpi(r) Y_{\l,m}(\theta,\phi). 
\Eeq
This decomposition automatically satisfies the divergence constraint.  The solutions for the toroidal and poloidal components decouple.  The radial component of the curl of \eq{eqn:diffusion_vel} gives
\Beq\label{eqn:toroidal}
(\Delta_\ell + \kappa^2) \mathcal{T}(r) = 0,
\Eeq
where
\Beq
\Delta_\ell = \frac{1}{r^2}\frac{d}{dr} r^2 \frac{d}{dr} - \frac{\ell(\ell+1)}{r^2}.
\Eeq
The horizontal divergence of \eq{eqn:diffusion_vel} gives an equation for the pressure,
\Beq
\varpi(r) = \frac{d}{dr}\left[r \left(\Delta_\ell + \kappa^2\right) \mathcal{P}(r)\right].
\Eeq
With this, the radial component of \eq{eqn:diffusion_vel} becomes
\Beq\label{eqn:poloidal}
\left(\Delta_\ell+k^2\right)\Delta_\ell \mathcal{P}(r) = 0.
\Eeq

The \eqs{eqn:toroidal}{eqn:poloidal} can be solved,
\Beq
\mathcal{T}(r) &=& A\, j_\ell(\kappa r), \\
\mathcal{P}(r) &=& B\, j_\ell(\kappa r) + C\, r^\ell, \label{Polo diffusion sol}
\Eeq
where $j_\ell(\kappa r)$ represents a spherical Bessel function of degree $\l$. The constants $A$, $B$, and $C$ must be chosen to satisfy the boundary conditions. The toroidal modes require one boundary condition to determine the ``dispersion relation'' for $\kappa$. 
The poloidal modes require two boundary condition; one to fix the coefficient of the  harmonic form $r^{\l}$, and the other the determine the dispersion relation. 

After substituting \eqss{vector-PT-solution}{Polo diffusion sol} into the boundary conditions found in \eqss{vec no-slip}{eqn:pseudo-vacuum} in \S4.3, we always find a dispersion relation of the form
\Beq
j_{\l + \mathrm{a}+2}(\kappa) \ = \ c_{\l}\, j_{\l+\mathrm{a}}(\kappa) \label{dispersion formula}
\Eeq
for some coefficients $c_{\l}$ and shift in the regularity class parameter, $\l + \mathrm{a}$. The spherical Bessel function are easy to evaluate with standard packages; \eg \url{scipy}. This make it simple to generate good guesses for the roots of \eq{dispersion formula}. We use Newton's method to  polish the zeros to a high degree of accuracy. Table~\ref{tab:geoff} shows the different dispersion (decay-rate) relations for different boundary conditions and modes. 

\begin{table*}
\label{dispersion table}
\centering
\begin{tabular}{lcccc}\hline
Boundary condition & \vline & Torordal & \vline & Poloidal  \\ \hline  \hline 

&\vline &  &\vline  & \\

no-slip & \vline & $j_{\l}(\kappa) $ & \vline &  $j_{\l+1}(\kappa)$ \\ 

&\vline &  &\vline  & \\
\hline
&\vline &  & \vline &\\

stress-free &\vline & $ j_{\l+1}(\kappa) - \frac{\l-1}{\l+2}\, j_{\l-1}(\kappa)$  & \vline & $j_{\l+2}(\kappa) - \frac{2\l+1}{2}\, j_{\l}(\kappa)$ \\

&\vline &  &\vline  & \\
\hline
&\vline &  &\vline & \\

potential-field & \vline& $j_{\l-1}(\kappa) $ & \vline & $j_{\l}(\kappa) $ \\

&\vline &  &\vline  & \\
\hline
&\vline &  &\vline & \\
 
perfect-conductor & \vline& $j_{\l}(\kappa) $ &\vline & $ j_{\l+1}(\kappa) - \frac{\l + 1}{\l}\, j_{\l-1}(\kappa)$ \\ 

&\vline &  &\vline  & \\
\hline
&\vline &  &\vline &\\

pseudo-vacum &\vline & $ j_{\l+1}(\kappa) - \frac{\l + 1}{\l}\, j_{\l-1}(\kappa)$ &\vline & $j_{\l}(\kappa) $ \\

&\vline &  &\vline &\\
\hline

\end{tabular}
\centering
\caption{The various dispersion formulae for different boundary conditions. In each case, the function is set to vanish and solved for $\kappa$ with Newton's method. The Toroidal and Poloidal modes each give different decay formulae for a give boundary condition.}\label{tab:geoff}
\end{table*}

\section{Matrices for Hydrodynamic Benchmark}\label{sec:matrices_hydro}

In our formulation of this problem, we use the statevector
\Beq
X \ = \ \left[
\begin{array}{c}
 u^- \\
 u^0 \\
 u^+ \\
 p
\end{array}
\right].
\Eeq
The linear operators $M$ and $L$ are
\Beq
M \ &=& \ \left[
\begin{array}{cccc}
 C_{1,\ell-1}C_{0,\ell-1} & 0 & 0 & 0 \\
 0 & C_{1,\ell}C_{0,\ell} & 0 & 0 \\
 0 & 0 & C_{1,\ell+1}C_{0,\ell+1} & 0 \\
 0 & 0 & 0 & 0
\end{array}
\right], \\
L \ &=& \ \left[
\begin{array}{cccc}
 -D_{1,\ell}^-D_{0,\ell-1}^+ & 0 & 0 & \xi_\ell^-C_{1,\ell-1}D_{0,\ell}^- \\
 0 & -D_{1,\ell+1}^-D_{0,\ell}^+ & 0 & 0 \\
 0 & 0 & -D_{1,\ell}^+D_{0,\ell+1}^- & \xi_{\ell}^+C_{1,\ell+1}D_{0,\ell}^+ \\
 \xi_{\ell}^-D_{0,\ell-1}^+ & 0 & \xi_{\ell}^+D_{0,\ell+1}^- & 0
\end{array}
\right].
\Eeq
Thus, the part of the problem treated implicitly is identical to the linear diffusion equation described in section~\ref{sec:diffusion}.  The explicit terms are
\Beq
F(X) \ = \ \left[
\begin{array}{c}
 C_{1,\ell-1}C_{0,\ell-1}\left(-\vec{u}\vec{\cdot}\vec{\nabla}\vec{u} - 2 \Omega \vec{e}_z\vec{\times}\vec{u}\right)^- \\
 C_{1,\ell}C_{0,\ell}\left(-\vec{u}\vec{\cdot}\vec{\nabla}\vec{u} - 2 \Omega \vec{e}_z\vec{\times}\vec{u}\right)^0 \\
 C_{1,\ell+1}C_{0,\ell+1}\left(-\vec{u}\vec{\cdot}\vec{\nabla}\vec{u} - 2 \Omega \vec{e}_z\vec{\times}\vec{u}\right)^+ \\
 0
\end{array}
\right].
\Eeq
For boundary conditions, we either replace the last rows of the three components of $\vec{u}$ in the $L$ and $M$ matrices with the $r=1$ operator, and then replace the corresponding entries of $F(X)$ with the appropriate components of $\vec{u}_0$ ($\alpha_{BC}=2$), or we impose the boundary conditions with $\tau$ corrections ($\alpha_{BC}=0$).  We fix the $\ell=0$ component of all fields to be zero.

\section{Matrices for Convection Benchmark}\label{sec:conv matrices}

We use the statevector
\Beq
X \ = \ \left[
\begin{array}{c}
 u^- \\
 u^0 \\
 u^+ \\
 p \\
 T
\end{array}
\right],
\Eeq
The linear operators $M$ and $L$ are
\Beq
M \ = \ \left[
\begin{array}{ccccc}
 E C_{1,\ell-1}C_{0,\ell-1} & 0 & 0 & 0 & 0 \\
 0 & E C_{1,\ell}C_{0,\ell} & 0 & 0 & 0 \\
 0 & 0 & E C_{1,\ell+1}C_{0,\ell+1} & 0 & 0 \\
 0 & 0 & 0 & 0 & 0 \\
 0 & 0 & 0 & 0 & \textit{Pr}\, C_{1,\ell}C_{0,\ell}
\end{array}
\right],
\Eeq
\Beq
L \ = \ \left[
\begin{array}{ccccc}
 -E D_{1,\ell}^-D_{0,\ell-1}^+ & 0 & 0 & \xi_\ell^-C_{1,\ell-1}D_{0,\ell}^- & 0 \\
 0 & -E D_{1,\ell+1}^-D_{0,\ell}^+ & 0 & 0 & 0 \\
 0 & 0 & - E D_{1,\ell}^+D_{0,\ell+1}^- & \xi_{\ell}^+C_{1,\ell+1}D_{0,\ell}^+ & 0 \\
 \xi_{\ell}^-D_{0,\ell-1}^+ & 0 & \xi_{\ell}^+D_{0,\ell+1}^- & 0 & 0 \\
 0 & 0 & 0 & 0 & -\textit{Pr}\, D_{1,\ell+1}^-D_{0,\ell}^+
\end{array}
\right].
\Eeq
The explicit terms are
\Beq
F(X) \ = \ \left[
\begin{array}{c}
 C_{1,\ell-1}C_{0,\ell-1}\left(-E \vec{u}\vec{\cdot}\vec{\nabla}\vec{u} - \vec{e}_z\vec{\times}\vec{u} + \textit{Ra}\, T\, r\vec{e}_r\right)^- \\
 C_{1,\ell}C_{0,\ell}\left(-E \vec{u}\vec{\cdot}\vec{\nabla}\vec{u} - \vec{e}_z\vec{\times}\vec{u} + \textit{Ra}\, T \, \vec{r}\right)^0 \\
 C_{1,\ell+1}C_{0,\ell+1}\left(-E \vec{u}\vec{\cdot}\vec{\nabla}\vec{u} - \vec{e}_z\vec{\times}\vec{u} + \textit{Ra}\, T \, \vec{r} \right)^+ \\
 0 \\
 C_{1,\ell}C_{0,\ell}\left(S - \textit{Pr}\, \vec{u}\vec{\cdot}\vec{\nabla} T\right)
\end{array}
\right].
\Eeq
For the boundary conditions, we replace the bottom row of the three velocity components and temperature blocks with the impenetrable, stress-free, and fixed temperature conditions (see section~\ref{sec:diffusion}) when we use $\alpha_{BC}=2$, and implement the boundary conditions using $\tau$ corrections for $\alpha_{BC}=0$.  For the $\ell=0$ mode, we set the pressure and velocities to zero, but evolve the temperature equation normally.

\section{Matrices for Dynamo Benchmark}\label{sec:dynamo matrices}

Our statevector is
\Beq
X \ = \ \left[
\begin{array}{c}
 u^- \\
 u^0 \\
 u^+ \\
 p \\
 T \\
 A^- \\
 A^0 \\
 A^+ \\
 \Phi
\end{array}
\right],
\Eeq
The linear operator $M$ is
\Beq
M \ = \ \left[
\begin{array}{ccccccccc}
 M_{00} & 0 & 0 & 0 & 0 & 0 & 0 & 0 & 0 \\
 0 & M_{11} & 0 & 0 & 0 & 0 & 0 & 0 & 0 \\
 0 & 0 & M_{22} & 0 & 0 & 0 & 0 & 0 & 0 \\
 0 & 0 & 0 & 0 & 0 & 0 & 0 & 0 & 0 \\
 0 & 0 & 0 & 0 & M_{44} & 0 & 0 & 0 & 0 \\
 0 & 0 & 0 & 0 & 0 & M_{55} & 0 & 0 & 0 \\
 0 & 0 & 0 & 0 & 0 & 0 & M_{66} & 0 & 0 \\
 0 & 0 & 0 & 0 & 0 & 0 & 0 & M_{77} & 0 \\
 0 & 0 & 0 & 0 & 0 & 0 & 0 & 0 & 0
\end{array}
\right],
\Eeq
where
\Beq
M_{00}=\textit{Ro}\,C_{1,\ell-1}C_{0,\ell-1}, \quad\quad M_{11}=\textit{Ro}\,C_{1,\ell} C_{0,\ell}, \quad \quad M_{22}&=&\textit{Ro}\,C_{1,\ell+1}C_{0,\ell+1}, \nonumber \\
M_{55}=C_{1,\ell-1}C_{0,\ell-1}, \quad\quad M_{44}=M_{66} = C_{1,\ell}C_{0,\ell}, \quad\quad M_{77}&=&C_{1,\ell+1}C_{0,\ell+1}. \nonumber
\Eeq
The linear operator $L$ is
\Beq
L \ = \ \left[
\begin{array}{ccccccccc}
 L_{00} & 0 & 0 & L_{03} & 0 & 0 & 0 & 0 & 0 \\
 0 & L_{11} & 0 & 0 & 0 & 0 & 0 & 0 & 0 \\
 0 & 0 & L_{22} & L_{23} & 0 & 0 & 0 & 0 & 0 \\
 L_{30} & 0 & L_{32} & 0 & 0 & 0 & 0 & 0 & 0 \\
 0 & 0 & 0 & 0 & L_{44} & 0 & 0 & 0 & 0 \\
 0 & 0 & 0 & 0 & 0 & L_{55} & 0 & 0 & L_{58} \\
 0 & 0 & 0 & 0 & 0 & 0 & L_{66} & 0 & 0 \\
 0 & 0 & 0 & 0 & 0 & 0 & 0 & L_{77} & L_{73}  \\
 0 & 0 & 0 & 0 & 0 & L_{85} & 0 & L_{87} & 0 
\end{array}
\right],
\Eeq
where
\Beq
\begin{array}{lclcl}
L_{00} = -E D_{1,\ell}^-D_{0,\ell-1}^+, & & L_{11} = -E D_{1,\ell+1}^-D_{0,\ell}^+, & & L_{22} = - E D_{1,\ell}^+D_{0,\ell+1}^-, \\
L_{44} = -q D_{1,\ell+1}^-D_{0,\ell}^+, & & L_{55} =  -D_{1,\ell}^-D_{0,\ell-1}^+, & & L_{66} = -D_{1,\ell+1}^-D_{0,\ell}^+, \\
L_{77} = - D_{1,\ell}^+D_{0,\ell+1}^- & & L_{03} = L_{58} =\xi_\ell^-C_{1,\ell-1}D_{0,\ell}^- & & L_{23} = L_{78} =\xi_{\ell}^+C_{1,\ell+1}D_{0,\ell}^+, \\
L_{30} = L_{85} = \xi_{\ell}^-D_{0,\ell-1}^+, & & L_{32} = L_{87} =\xi_{\ell}^+D_{0,\ell+1}^-. & &
\end{array}
\Eeq
The explicit terms are
\Beq
F(X) \ = \ \left[
\begin{array}{c}
 C_{1,\ell-1}C_{0,\ell-1}\left(-\textit{Ro} \vec{u}\vec{\cdot}\vec{\nabla}\vec{u} - \vec{e}_z\vec{\times}\vec{u} + q\, \textit{Ra}\, T \, \vec{r} + \vec{B}\vec{\cdot}\vec{\nabla}\vec{B}\right)^- \\
 C_{1,\ell}C_{0,\ell}\left(-\textit{Ro} \vec{u}\vec{\cdot}\vec{\nabla}\vec{u} - \vec{e}_z\vec{\times}\vec{u} + q\, \textit{Ra}\, T \, \vec{r} + \vec{B}\vec{\cdot}\vec{\nabla}\vec{B}\right)^0 \\
 C_{1,\ell+1}C_{0,\ell+1}\left(-\textit{Ro} \vec{u}\vec{\cdot}\vec{\nabla}\vec{u} - \vec{e}_z\vec{\times}\vec{u} + q\,\textit{Ra}\, T \,  \vec{r} + \vec{B}\vec{\cdot}\vec{\nabla}\vec{B}\right)^+ \\
 0 \\
 C_{1,\ell}C_{0,\ell}\left(S - \vec{u}\vec{\cdot}\vec{\nabla} T\right) \\
 C_{1,\ell-1} C_{0,\ell-1}\left(\vec{u}\vec{\times}\vec{B}\right)^- \\
 C_{1,\ell}C_{0,\ell}\left(\vec{u}\vec{\times}\vec{B}\right)^0 \\
 C_{1,\ell+1} C_{0,\ell+1}\left(\vec{u}\vec{\times}\vec{B}\right)^+ \\
 0
\end{array}
\right].
\Eeq
For the boundary conditions, we replace the bottom row of the three velocity components, three magnetic vector potential components and temperature blocks with the impenetrable, stress-free, potential, and fixed temperature conditions (see section~\ref{sec:diffusion}) for $\alpha_{BC}=2$, or impose the boundary conditions using $\tau$ corrections for $\alpha_{BC}=0$.  For the $\ell=0$ mode, we set $p$, $\Phi$, velocities and magnetic vector potential to zero, but evolve the temperature equation normally.

\end{document}